**Shape-Determined Kinetic Pathways in 2D Solid-Solid Phase Transitions**

*Ruijian Zhu, Yi Peng, Yanting Wang**

R. Zhu, Y. Wang
Institute of Theoretical Physics, Chinese Academy of Sciences, Beijing 100190, China
E-mail: wangyt@itp.ac.cn

Y. Peng
Beijing National Laboratory for Condensed Matter Physics, Institute of Physics, Chinese Academy of Sciences, Beijing 100190, China

R. Zhu, Y. Peng, Y. Wang
School of Physical Sciences, University of Chinese Academy of Sciences, 19A Yuquan Road, Beijing 100049, China

Funding: National Natural Science Foundation of China (No.12047503)

## Abstract

Solid-solid phase transitions are ubiquitous in nature, but the kinetic pathway of anisotropic particle systems remains elusive, where the coupling between translational and rotational motions plays a critical role in various kinetic processes. Here we investigate this problem by molecular dynamics simulation for two-dimensional ball-stick polygon systems, where pentagon, hexagon, and octagon systems all undergo an isostructural solid-solid phase transition. During heating, the translational motion exhibits merely a homogeneous expansion, whereas the time evolution of body-orientation is shape-determined. The local defects of body-orientation self-organize into a vague stripe for pentagon, a random pattern for hexagon, while a distinct stripe for octagon. The underlying kinetic pathway of octagon adheres to the quasi-equilibrium assumption, whereas the pathways of hexagon and pentagon are governed by translational and rotational motion, respectively. This diversity is originated from different kinetic coupling modes determined by the anisotropy of molecules, and can affect the phase transition rates. The reverse process in terms of cooling follows the same mechanism, with more diverse kinetic pathways attributed to the possible kinetic traps. Our findings promote the theoretical understanding of microscopic kinetics of solid-solid phase transitions as well as provide direct guidance for the rational design of materials utilizing desired kinetic features.

## 1. Introduction

Solid-solid (s-s) phase transition is one of the most ubiquitous types of phase transitions in natural and man-made materials, including alloys[1], minerals[2], ionic liquids[3], ice[4,5] and biological processes[6]. S-s transitions also hold considerable technological importance in earth science[7,8], diamond production[9], memory alloys[10-12], reconfigurable optical devices[13] and reprogramming self-assembly[14-17]. Specifically, 2D molecular crystals stabilized by non-covalent interaction have become a hot topic in materials science and chemical engineering[18-20], where the solid-solid transition has already been observed and is suggested to be useful for molecular electronic devices[21]. Beyond thermodynamic aspect, it has been shown that different kinetic pathways can influence both phase transition rates[22-24] and final products[23,25] in the s-s transitions. The kinetic pathway follows the free energy landscape if the quasi-equilibrium assumption is satisfied, but can also deviate when kinetic details are non-negligible[26].

In order to investigate the microscopic kinetics of s-s transitions, spherical colloids, benefiting from the single-particle-resolution and the controllability of interactions and sizes[27], serve as outstanding model systems. To date, extensive studies have revealed various kinetic pathways modulated by external stress[24,28,29], electric field[23], grain boundaries[24,30,31], and internal softness[30]. However, spherical colloids are not suitable for studying kinetic pathways of s-s transitions in anisotropic systems, where monomer shape critically impacts the transition pathway via kinetic coupling between translational and rotational motions, as demonstrated in crystallization[32-35], vitrification[36], and the formation of polycrystals[37]. Although s-s transitions have also been observed in a variety of three-dimensional (3D)[22,38-40] and two-dimensional (2D)[35,41-48] systems composed of anisotropic particles through molecular simulations or colloidal experiments, prior works predominantly focus on either thermodynamics or kinetics of translational motion only.

Our recent theoretical study[48] on 2D ball-stick polygon systems, composed of one L-J ball on each vertex and valence bonds between adjacent balls, has revealed complex phase behaviors despite the simplicity of the model. Specifically, the ball-stick pentagon, hexagon, and octagon all exhibit s-s transitions from the close-packing phase to the rotator crystal phase during heating at an appropriate pressure, whose structures are shown in **Figure 1**(a)-(c). These s-s transitions are isostructural[49,50], as observed in some colloidal systems[51,52] and colloid-polymer mixtures[53], where the lattice homogeneously expands at the phase transition point drastically with its symmetry group retained. Given that their translational motions are well-understood, these ball-stick polygons are ideal for exploring how the coupling between translational and rotational motions affects the kinetic pathways of s-s transitions in anisotropic particle systems.

In this work, we investigate the kinetics of the s-s phase transitions of ball-stick pentagon, hexagon, and octagon systems by molecular dynamics (MD) simulation. During heating, the amorphization of the well-ordered body-orientations in the parent phase occurs concurrently with the homogeneous expansion. The rotational motion is quantified by the local defects of the body-orientation, which arrange into three qualitatively different patterns: The defects for pentagon form a vague stripe, the ones for hexagon create a random pattern, whereas the ones for octagon produce a distinct stripe. Our mechanistic analysis reveals shape-determined kinetic pathways originated from different kinetic coupling modes: pentagon has a kinetic pathway dominated by rotational motion, hexagon is dominated by translational motion, and octagon satisfies the quasi-equilibrium assumption where the rotational and translational motions are synchronous. Different kinetic pathways can directly influence the phase transition rate. The reverse process follows the same mechanism with more diverse kinetic pathways due to kinetic traps. By examining the local structures of the crystalline states for different polygons, we conclude that a stronger rotational constraint promotes a kinetic pathway preferring translational motion and favoring rotational motion in the reverse process, and *vice versa*.

## 2. Results

### 2.1. Macroscopic Features of S-S Transitions

As shown in Figure 1(a)-(c), the ground state of the ball-stick pentagon system adopts a striped phase[48], where monomers are aligned alternatively along two parallel body-orientations, and their center-of-masses (COMs) form a triangular-lattice structure; for both hexagon and octagon, the ground states are triangular-lattice with uniform body-orientation. The balls form a structure closed to distorted honeycomb lattice in the hexagon system, but no clear lattice structures formed in pentagon and octagon systems. Throughout this paper, the term *monomer* refers to one ball-stick polygon and only the lattice structures of COMs are concerned. Under an appropriate temperature, the lattice undergoes a sudden expansion, leading to a sharp increase in potential energy and a concurrent decrease in density, which are typical features for a discontinuous phase transition. This rise in potential energy is compensated by the larger entropy gained from the loss of body-orientational order, resulting in a rotator crystal phase that preserves only translational order, evidenced by a more uniform and broader distribution of body-orientation angle[48]. Both the parent and product phases in these three systems can be characterized by the bond-orientational order parameter $\Psi_6$ [48], manifesting a six-fold symmetry of the lattices. The same symmetry between parent and product lattices confirms the isostructural nature of these s-s transitions[49,50]. Despite the similar thermodynamic behavior, it would be interesting to know whether the kinetic pathways of the s-s transitions for different polygons exhibit shape-dependent features.

**Figure 1. Crystalline morphologies and rotational motions of the s-s transitions in three ball-stick polygon systems.** (a) Crystal states of pentagon. (b) Crystal states of hexagon. (c) Crystal states of octagon. In (a)-(c), the upper panels are the snapshots of ground-state morphologies with monomers colored in blue, sharing an ordered body-orientation, and the lower panels are the snapshots of rotator crystal state with monomers colored in orange, exhibiting random body-orientations. (d)-(f) show the spatial distributions of local defects in the body-orientation fields with warmer colors indicating a higher defect occurrence rate. The points with high occurrence rates form a vague striped region along with some randomly distributed points in the pentagon system (d), exhibit a highly random pattern in the hexagon system (e), and demonstrate a distinct striped region across the box in the octagon system (f). The results showed here are all obtained at the lowest investigated pressure of the corresponding polygon ($P$ = 0 for pentagon and octagon while $P$ = 1.5 for hexagon).

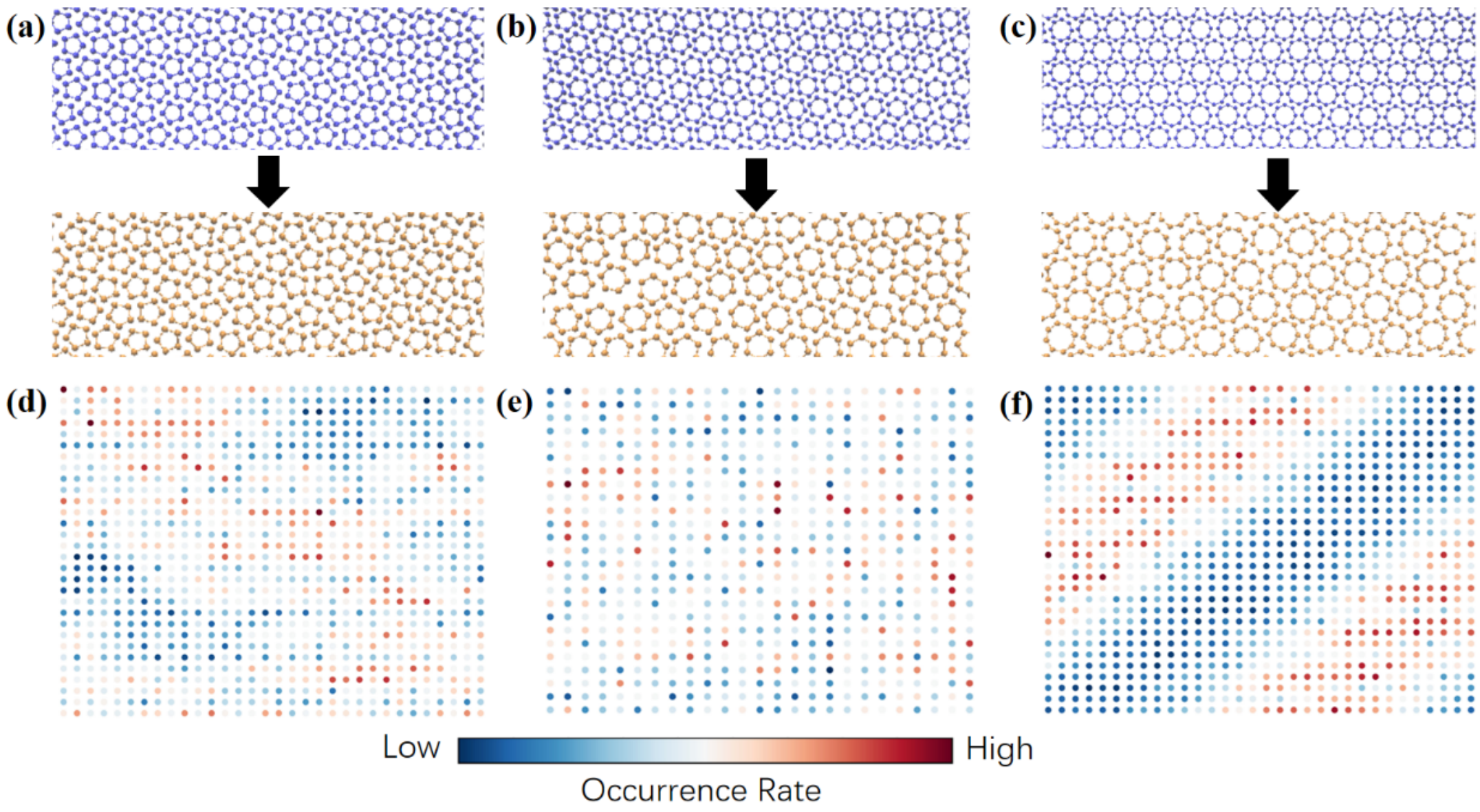

Before exploring the microscopic details of kinetics, let us first take a look at the general features of translational and rotational motions. Starting from the ground-state crystalline morphology equilibrated at a temperature slightly below the phase transition point, the s-s transition is triggered by heating the system across the phase transition point. During the transition, the lattice undergoes a homogeneous spatial expansion, quantified by the time evolution of Voronoi cell areas associated with the lattice points, as shown in Figure S1(a)-(c). The area distribution retains a unimodal shape throughout the transition, with its peak continuously shifting toward the direction corresponding to a larger area. This behavior is observed in all the three polygons, consistent with the previous theoretical prediction[49,50]. The inter-monomer distances of the expanding lattice increase, leading to a greater free volume for the particle to rotate. Consequently, once the density of the system starts decreasing, the body-orientational parameter defined as $\phi_i = \exp(in\theta_i)$ also begins to decrease, as can be seen in Figure S2. Here $\theta_i$ is the angle between an arbitrary vertex of polygon $i$ and the $x$-axis, and $n$ takes the $n$-fold symmetry of a polygon into account. For hexagon and octagon, $n$ is just the edge number, whereas for pentagon, $n$ is set to 10 because there are two alternative body-orientations in the ground state. When the body-orientation of the monomers becomes disordered, the lattice takes several more steps to relax to its final structure, forming a rotator crystal.

## 2.2. Defects in the Body-Orientation Field

More detailed features of the rotational motion are obtained by examining the *body-orientation field*, defined as a collection of unit vectors located at the monomer COMs, whose directions represent the arguments of the body-orientational order parameters of the monomers with respect to the $x$-axis. All vectors are roughly aligned along the same direction in the parent phase, and become disordered in the product phase. To quantify this amorphization process, we track the production and annihilation of local defects (defined in the Methods section), which remark the positions where body-orientations change drastically.

For the pentagon system, some defects already present in the parent phase, as shown in Figure S3(a) in SI. As the transition proceeds, the number of defects increases significantly, which initially form a striped region across the simulation box and eventually become uniformly distributed, as shown in Figure S3(b)-(f). This feature is quantified by the spatial distribution of local defects (detailed in the Methods section). By averaging over all the snapshots sampled during the transition, a vague striped region exhibits along the crystal axis direction across the box, as shown in Figure 1(d), alongside isolated defects wherever easy to create without significant expansion.

For the hexagon system, very few defects are present in the parent phase, as shown in Figure S4(a). As the transition progresses, defects begin to appear randomly in the box, as shown in Figure S4(b)-(d). Unlike the pentagon system, no regular patterns emerge during the entire transition process. The spatial distribution of defects shown in Figure 1(e) is totally random, coinciding with the snapshots shown in Figure S4.

During the transition process, the octagon system initially behaves similarly to the hexagon system and later resembles the behavior of the pentagon system. A very small number of randomly distributed defects initially appear and subsequently disappear due to thermal fluctuations before the formation of a stable narrow stripe across the box. The striped region then extends along its normal direction via a rate-limited process, until finally uniformly covers the whole simulation box. The typical snapshots demonstrating the above process are presented in Figure S5. The distinct stripe across the box for the spatial distribution of local defects appeared in Figure 1(f) supports the above physical picture.

### 2.3. Diversity of Kinetic Pathways

The above three observed patterns imply a shape-determined nature of the kinetic pathway. To reveal the underlying mechanism in terms of kinetic coupling modes of translational and rotational motions, it is instructive to fix the translational motion while allow the monomers to rotate freely in MD simulation (hereafter referred as 'fixed simulation'). For the fixed simulations, we randomly select several conformations along the transition process in the original trajectory to serve as initial configurations with the COM position of each monomer being fixed, and then perform MD simulations in the *NVT* ensemble (more details can be found in the Methods section and SI). Since the lattice homogeneously expands, each snapshot can represent the COM positions at a given volume. Therefore, this constrained system spontaneously relaxes to the equilibrium state given by the restricted free energy $F(V,\Phi)$, where the ensemble average of body-orientational order parameter $\Phi$ directly corresponds to the minimum satisfying $(\frac{\partial F(V,\Phi)}{\partial \Phi})_V = 0$. Below we present the ensemble-averaged results of $\Phi$ and compare them with the values obtained from the unfixed simulation, i.e., the original MD simulation, with the same volume. More information regarding the time evolution of the body-orientation and the selection of initial configurations for the fixed simulations can be found in SI.

The fixed simulations have revealed three qualitatively different kinetic pathways for the polygon systems: (1) For pentagon, as shown in **Figure 2**(a), the fixed simulations give slightly larger or equal equilibrium values of $\Phi$ than the unfixed simulation, indicating that the rotational motion is marginally faster than the translational motion, i.e., the kinetic process is led by rotation; (2) By contrast, for hexagon, the fixed simulation always produce a smaller equilibrium value of $\Phi$ than the unfixed simulation, as shown in Figure 2(b), suggesting that the translational motion is faster and the transition process is dominated by expansion; (3) The case for octagon shown in Figure 2(c) falls in between, where the equilibrium values of $\Phi$ from the fixed simulations are consistently identical with the unfixed simulations, indicating that the rotational and translational motions are synchronous in the transition process.

**Figure 2. The results from the fixed simulations.** (a) Pentagon. The equilibrium values from the fixed simulations are slightly larger than or equal to the values from the unfixed MD

simulation. (b) Hexagon. The equilibrium values from the fixed simulations are consistently and significantly smaller than the values from the unfixed MD simulation. (c) Octagon. The equilibrium values from the fix simulations are always equal to the values from the unfixed MD simulation. The inset of each subfigure illustrates that, in the fixed simulations, each monomer has its COM fixed but is totally free to rotate. In (a)-(c), each data point labeled ‘fixed’ comes from an individual fixed simulation, and each point labeled ‘unfixed’ comes from a single trajectory of a regular MD simulation by heating up the parent phase. (d) Schematic illustration of the real kinetic pathways and the pathway satisfying the quasi-equilibrium assumption.

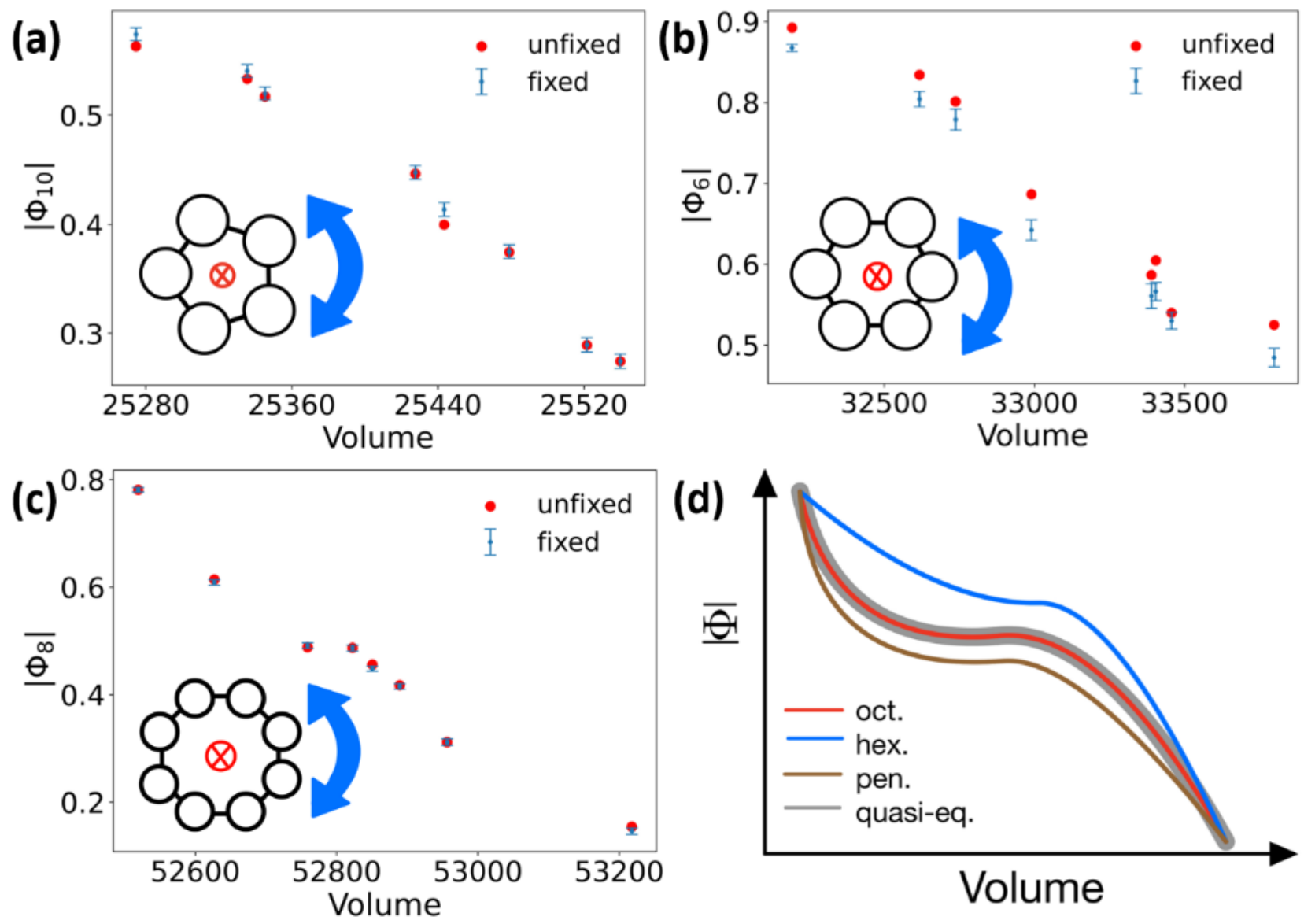

The kinetic pathways of s-s transition are widely regarded as following the thermodynamic free energy landscape[22,54], which however, makes sense only if the quasi-equilibrium assumption is satisfied[26]. This assumption requires that the kinetic pathway follows exactly the one connecting the minima of the constrained free energy values at different volumes, i.e., the pathway connecting the points satisfying $\frac{\partial F(V,\Phi)}{\partial \Phi}=0$ for all possible volumes $V$. As schematically illustrated in Figure 2(d), the kinetic pathway of pentagon is featured by a faster evolution of body-orientation than in the quasi-equilibrium pathway given by the thermodynamic free energy landscape, the one of hexagon suggests a faster evolution of volume, while only the one of octagon follows the quasi-equilibrium pathway. The violation of the quasi-equilibrium assumption suggests that the details of kinetics are non-negligible for the transition process[26]. Furthermore, it should be noted that the kinetic pathway dominated by rotation is close to the quasi-equilibrium one, because the rotational motion is significantly retarded by the repulsion of neighboring monomers; on the other hand, expansion faces no explicit obstacles, allowing the pathway dominated by the translational motion to be far from the quasi-equilibrium one.

The identified kinetic pathways can effectively explain the patterns of defects shown in Figure 1(d)-(f). In the hexagon system, where the expansion dominates the kinetic process, the relatively large distance between adjacent monomers allows the rotational motion to occur independently, resulting in a random pattern of local defects. By contrast, as for the case of pentagon and octagon, the neighboring monomers hinder the rotational motion due to the small interparticle distance. Consequently, when one particle changes its body-orientation due to thermal fluctuation, it is more likely to induce a rotation for its neighboring monomers rather than those located far away. This is analogous to the fact that the second low-energy excitation state of 2D Ising model is flipping two adjacent spins instead of two separated ones[55]. The defects then propagate along one of the crystalline axes, forming a narrow striped region, which resembles a typical configuration of phase coexistence in 2D system with periodic boundary condition[56]. The difference between pentagon and octagon is that the rotational motion is relatively easy in the pentagon system, indicated by the existence of a small number of defects in the parent phase, which results in some isolated defects besides the main striped region.
Although the above results are all obtained at the lowest pressures ($P = 0$ for both pentagon and octagon, while $P = 1.5$ for hexagon, more discussions can be found in the Methods section) for the s-s transitions, more simulations at various pressures have shown that both the kinetic pathways and the patterns of local defects in the body-orientation field are robust with respect to the pressure, indicating that they only rely on intrinsic properties of monomers. The simulation results at other pressures can be found in SI.

### 2.4. Phase Transition Rate

The diversity of kinetic pathways directly affects the phase transition rate, which is quantified in this work by the simulation time required to finish the phase transition (detailed in the Methods section). The starting point of the transition is selected as the moment when the potential energy of the system significantly deviates from the average value in the parent phase without reverting, and the endpoint is chosen as the time when the potential energy of the system reaches the average value in the product phase without obvious decrease thereafter. Although the choices of the starting and ending points are not very rigorous, they should not affect the qualitative conclusions we draw. As depicted in **Figure 3**, hexagon exhibits a relatively stable phase transition rate across various pressures, whereas both pentagon and octagon show a significantly accelerated transition rate as the pressure increases. Combining with the thermodynamic data shown in Table. S1, it can be inferred that the energy barrier is lower at a higher pressure due to reduced structural difference between the parent and product phases. For pentagon and octagon, where the rotational motion is pivotal in the transition process, the faster transition rate benefits from the lower energy barrier. This is also visible from the body-orientation field shown in Figure S6, Figure S8, Figure S9, and Figure S11, where the time required to form a narrow stripe is nearly constant at various pressures, but the growth of the stripe along its normal direction is faster at higher pressures. In the case of hexagon, the kinetic process is dominated by expansion, which has little to do with the energy barrier retarding rotational motion. Since the increases in phase-transition temperature and pressure balance each other, the hexagon system has a relatively stable speed for the translational motion, and thus the overall transition rate.

**Figure 3. Phase transition times for various polygons.** Each data point is averaged over ten independent simulation trajectories. Five of the ten trajectories are simulated at $T_m$, while the other five are simulated at $T_m + 0.01$. The phase transition time is expressed in the unit of $\tau \equiv \sqrt{m\sigma^2 / \varepsilon}$ . The $y$-axis adopts a logarithmic scale. The error-bars represent the standard deviation, resulting from the randomness among different simulation trajectories.

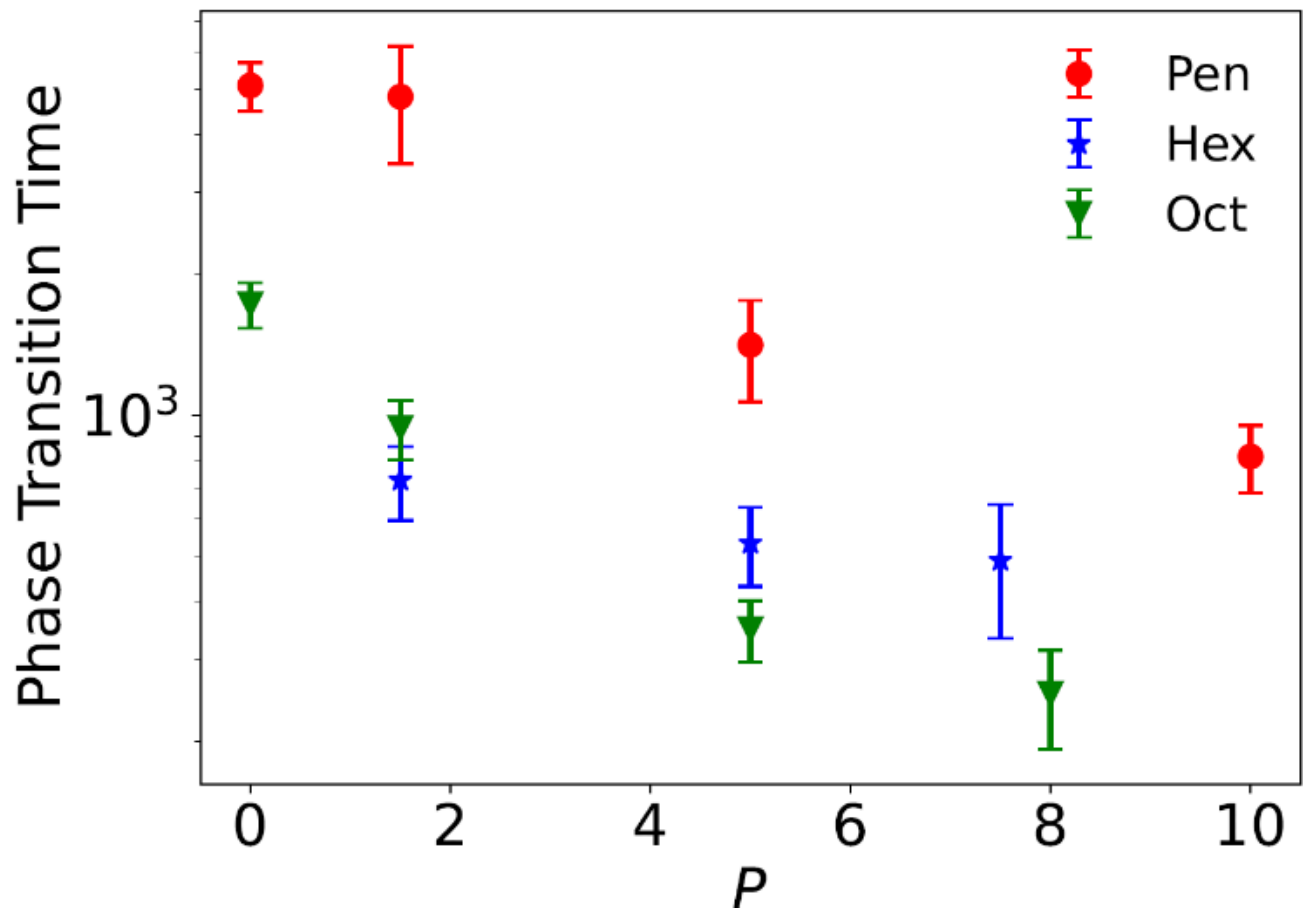

### 2.5. Reverse Process

The reverse process is studied by cooling down the rotator crystal state, during which the area of each Voronoi cell associated with the lattice points still retains a unimodal distribution, in agreement with the theoretical picture of homogeneous contraction[49,50], as shown in Figure S1 (d)-(f). The overall results for the three polygon systems are summarized in **Figure 4**(a). It is quite interesting that the reverse process for pentagon and octagon may approach either the corresponding ground state, as shown in Figure 4(b) and (d), or a metastable polycrystalline state formed by two large crystal fragments with different crystalline axes due to the kinetic trap. Both fragments adopt the ground-state crystalline structure, as depicted in Figure 4(e) for pentagon and (f) for octagon. Conversely, all six of our reverse simulations for hexagon consistently result in the ground state shown in Figure 4 (c). Since it is possible to reach a monocrystalline structure via cooling, it offers a route for recovering monocrystalline 2D materials from polycrystals by first heating it across the s-s transition temperature and then cooling it down.

To understand the effect of kinetic coupling in the reverse process, we also perform the fixed simulations for each polygon. The results suggest that the same polygon can have different kinetic pathways, as demonstrated in Figure S14. Despite the enhanced randomness compared to the heating process, the coupling mode has a strong correlation with the final product: When the kinetic process is dominated by contraction, the final product is always polycrystal; when it is dominated by rotation or satisfies the quasi-equilibrium assumption, the product forms a perfect crystal. In the scenarios where the rotational motion is faster, the monomers can effectively adjust their orientations. However, when translational motion prevails, the system is compressed into a high-density state, resulting in a kinetic trap devoid of global alignment of the monomers.

**Figure 4. Final structures of the reverse processes.** (a) Summary of the result that hexagon always ends up with the ground-state morphology (perfect lattice) while pentagon and octagon can reach either the perfect lattice or a polycrystal. (b) Perfect lattice of pentagon. (c) Perfect lattice of hexagon. (d) Perfect lattice of octagon. (e) Polycrystalline morphology of pentagon. (f) Polycrystalline morphology of octagon. The polycrystals basically consist of two pieces of perfect crystals with different orientations. In (b)-(f), each Voronoi cell associated with the lattice points is colored according to the body-orientation angle $\theta$ of corresponding monomer, ranging from 0 to $2\pi/n$.

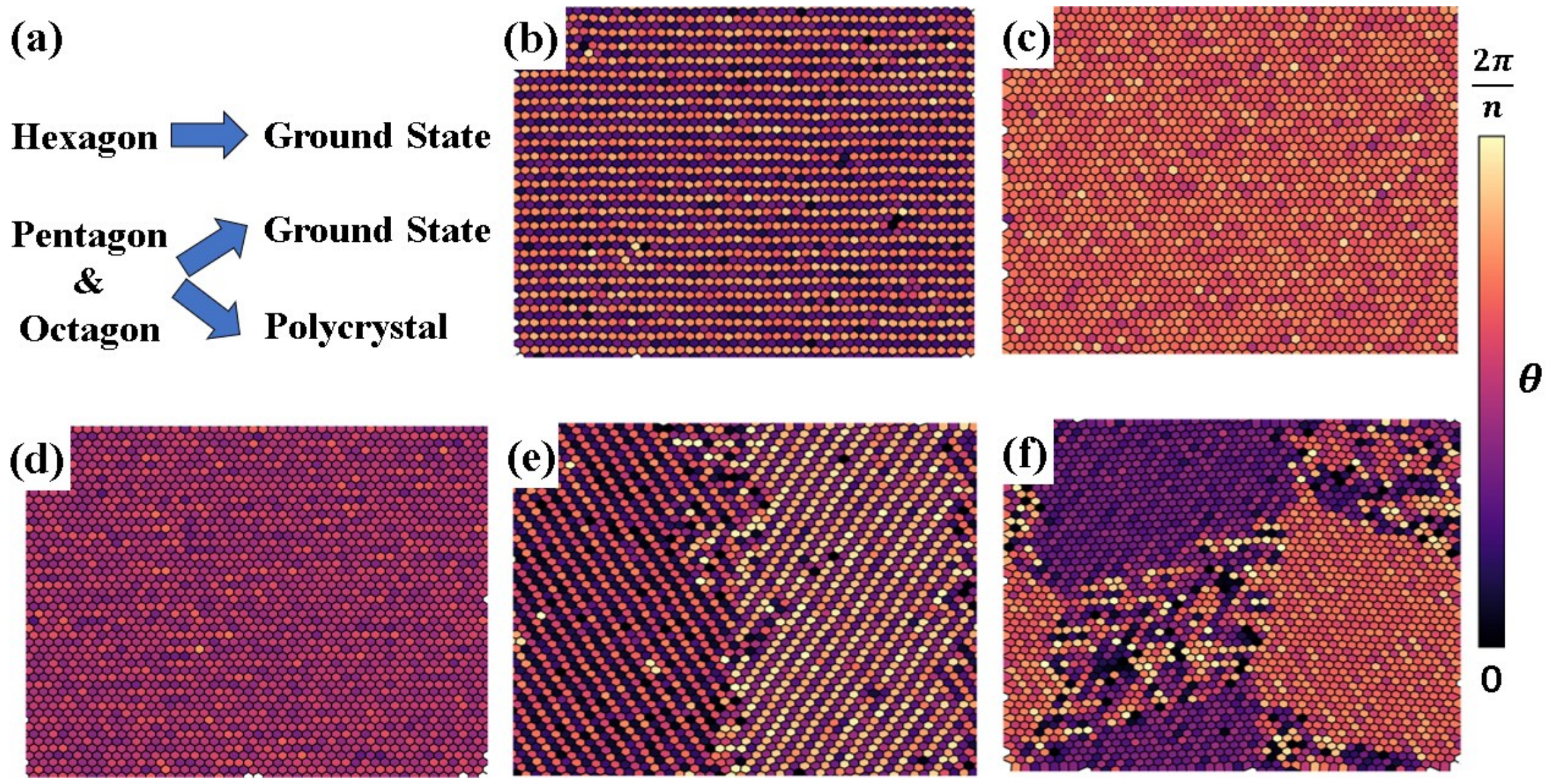

## 2.6. Connection between Kinetic Behaviors and Monomer Properties

In order to guide material design for targeted kinetic properties in solid-solid transitions, it is essential to relate the observed kinetic behaviors to molecular properties in terms of shape and interaction. These molecular properties govern the local structure in the close-packing phase, which in turn influences the phase transition kinetics. We quantify this local structure through statistical analyses of the relative distance $r_{12}$ and the relative body-orientation angle $\theta_{12}$, which is defined as the absolute value of the difference in body-orientation $\theta$ between two monomers, as illustrated in **Figure 5**(a). The absolute value guarantees a larger $\theta_{12}$ value simply corresponds to a larger deviation of the relative body-orientation between neighboring monomers. The *n*–fold symmetry is taking into account by limiting the value of $\theta$ in the range of 0 to $2\pi/n$, where *n* is the edge number of each polygon. As depicted in Figure 5(b)–(d), the ground state of pentagon exhibits a highly dispersive distribution of $\theta_{12}$, hexagon displays several distinct peaks with small dispersion, while octagon also suggests a centralized distribution, but approximately half of the neighbors of a monomer significantly deviate from its orientation. Because a stronger rotational constraint implies that the monomer in the crystalline state is difficult to change its body-orientation relative to its neighbors, these results indicate that pentagon has the weakest rotational constraint on neighboring monomers, octagon has an intermediate constraint, and hexagon has the strongest one.

A qualitative understanding on the rotational constraint can be obtained from the consideration on simply the shape of the monomer. Polygon with less edges exhibits a stronger orientational entropy[57], so it is expected that hexagon has a stronger rotational constraint than octagon. The pentagon, though with less edges than hexagon and octagon, has interlaced body-orientations of monomers in the striped phase, which significantly weakens the constraint on adjacent monomers with opposite orientations.

**Figure 5. Probability density maps of relative angle ($\theta_{12}$) and relative distance ($r_{12}$) between two monomers.** (a) Schematic illustration of the definitions of $\theta_{12}$ and $r_{12}$. (b) - (d)

are calculated in the close-packing phase and (e) - (g) are calculated in the rotator crystal phase. (b) and (e): pentagon. (c) and (f): hexagon. (d) and (g): octagon. Since the range of $\theta_{12}$ is different for different polygons, the probability density is normalized by the case of the uniform distribution.

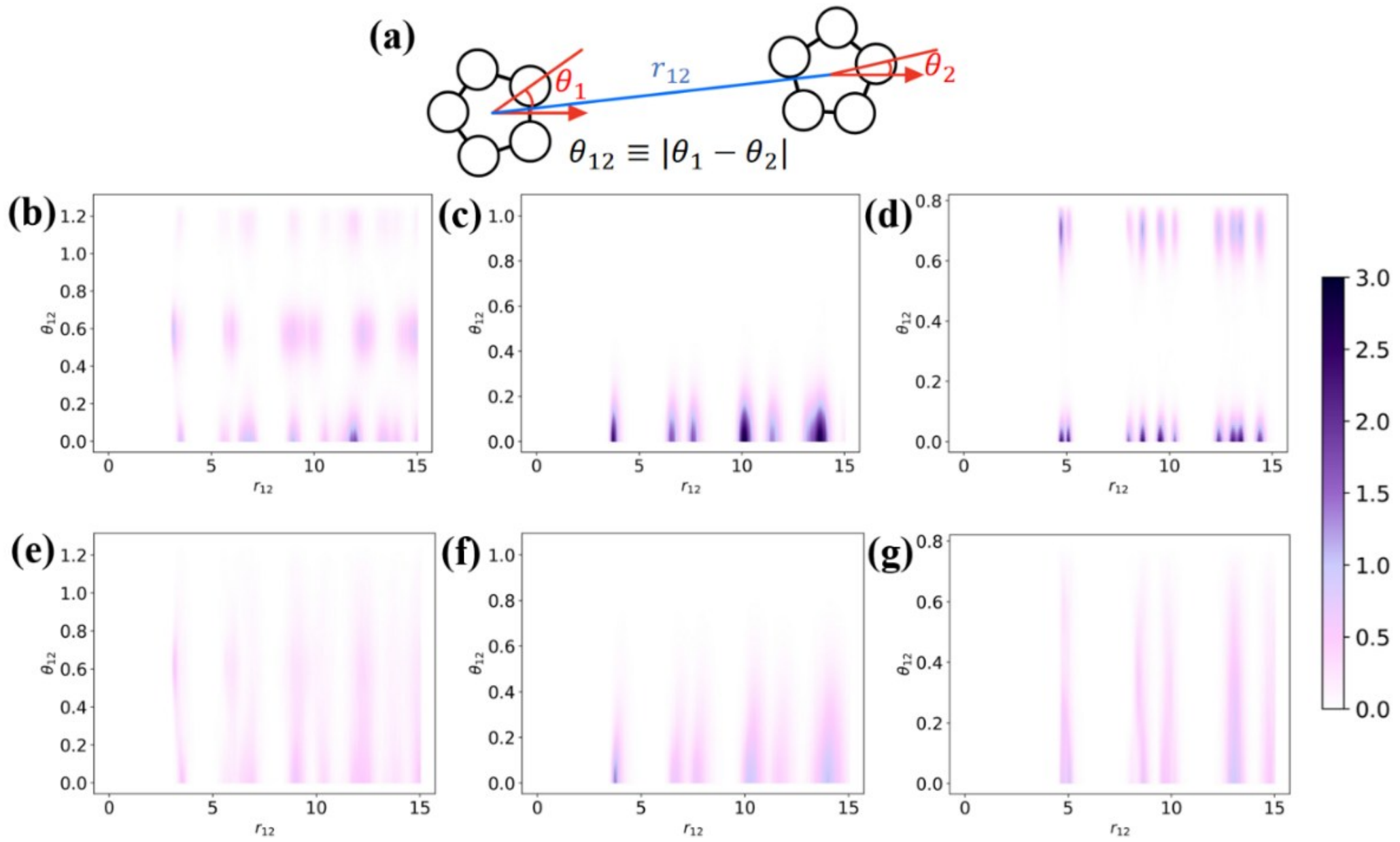

Because the interaction strength generally decays with the distance between monomers, when a strong rotational constraint is present, the system tends to expand before rotation to effectively attenuate the constraint. Conversely, with a weak constraint, the polygon can rotate without significant expansion.

Performing the same statistical analysis in the rotator crystal phase can also identify a direct connection between rotational constraints and kinetic behaviors. As shown in Figure 5(e)–(g), pentagon and octagon almost have no constraints on the relative angle of neighboring monomers, so the adjustment of the body-orientation cannot take place before contraction. As a result, the monomers may not align globally before contracted into a dense state. The hexagon, however, retains a moderate constraint on the relative angle, ensuring adequate modification of body-orientation before significant contraction.

## 3. Conclusions

In this work, we have revealed diverse kinetic pathways of s-s phase transitions in the 2D ball-stick polygons (pentagon, hexagon, and octagon) by MD simulation. During the transition from close-packing state to rotator crystal state, we have identified three different patterns for the time evolution of the body-orientation field, originated from the different kinetic-coupling modes between translational and rotational motions. The underlying kinetic pathway is shape-determined, either being dominated by one of the motions or satisfying the quasi-equilibrium assumption. One of the direct consequences is that the phase-transition rate is roughly constant at various pressures if translational motion is dominant, and otherwise it becomes faster at higher pressures. In the reverse process of cooling down the rotator crystal phase, the products of pentagon and octagon are either perfect crystals or polycrystals, whereas the product of hexagon is consistently the perfect crystalline structure. The final structure is strongly correlated with the kinetic pathway of this transition process. All the kinetic behaviors can be linked to the monomer properties in terms of the rotational constraints.

Despite the extensive studies on translational motion, the literature is less focused on the kinetic coupling between translational and rotational motions in s-s transitions, which is prevalent in almost all anisotropic particle systems and known to be crucial for various kinetic processes. Consequently, there is a gap between theoretical studies and novel s-s transitions in real material systems. Benefiting from the simplicity of the translational motion of the isostructural transitions, ball-stick polygon systems, exhibiting transitions from close-packing state to rotator crystal state, provide us an ideal platform for examining the effect of the kinetic coupling. The diversity of kinetic pathways, especially the ones violating the quasi-equilibrium assumption, emphasizes the importance of considering detailed kinetics, rather than merely concentrating on the thermodynamic free energy landscape. It has been shown that monomers ranging from atomic scale to macromolecules have similar kinetic and thermodynamic behaviors only if they share the same symmetry of shape and interaction[58]. Our results, which have been directly connected to monomer properties, are expected to be valid in a variety of realistic systems composed of 2D polygonal components, e.g., benzene, coronene, circumcoronenes[59,60], kekulene[61], and patchy colloids with designed symmetry[62]. Therefore, this work not only enhances theoretical understanding of the microscopic kinetics of the s-s transitions, but also provides guidance for the rational design of molecular devices with desired kinetic properties.

There are still some limitations in this work. For instance, in chemical synthesis or cell tissue[63,64], system usually shows polydisperse shape, and therefore a polygon mixture[65] should be considered. Moreover, the mechanism we propose here may be extended to s-s transitions with more complicated translational motion, e.g., the four-patch colloid with a set of inter-patch angles of (90°, 90°, 90°, 90°) exhibiting a transition from a square lattice to a triangular lattice[66] can be a good candidate. Lastly, it would be interesting to consider a quasi-2D system with large fluctuations on the *z*-direction. These problems can be investigated in the future to deepen our understanding on the effect of kinetic coupling and advance the applications in more realistic cases.

## 4. Methods

*Preparation of Initial Configurations*: Each ball-stick *n*-gon is composed of *n* L-J balls and *n* covalent bonds connecting adjacent balls. The covalent bond is represented by a harmonic potential with a very large strength ($k$ = 9000), retaining a regular polygon shape in the simulation, assisted by a harmonic potential with the same strength applied to each angle. For each polygon at a given pressure, we performed a regular MD simulation for $1.2\times10^7$ steps starting from the ground-state crystalline morphology at a temperature that is 0.01 lower than the transition point, and then the last snapshot of the simulation is used as the initial configuration for the heating process. The method used to determine the phase transition point has been described in ref. 44 and the related data can be found in the corresponding data repository[67].

Due to the hysteresis of first-order phase transition, the transition point determined by heating is higher than the one determined by cooling. Therefore, we prepared two kinds of initial configurations for the reverse process (cooling process). The first kind of initial configurations are prepared by equilibrating the rotator crystal phase obtained from the s-s transition described above, and the second ones are prepared by cooling down the rotator crystal phase to a temperature 0.01 higher than the phase transition point of the reverse process, and then relaxed for $1.2\times10^7$ steps. Our simulations indicate that the results of the reverse process do not depend on initial configurations.

*MD Simulations*: All the simulations in this work were conducted with LAMMPS[68]. Starting from the equilibrated structures of the parent phase, we performed the simulations in the isothermal-isotension ensemble[69] at $T_m$ and $T_m + 0.01$ with the periodic boundary conditions applied to both $x$ and $y$ directions, where $T_m$ is the phase transition point. The s-s transition process can be observed within $1.6\times10^7$ steps, evidenced by the increase of potential energy and the decrease of density. To determine the phase transition rate, we further performed ten independent simulations initiating from four different configurations (the last or the penultimate snapshots of the simulations at a temperature that is 0.01 or 0.02 lower than the phase transition point), and evaluated the standard error under each thermodynamic condition. Five of these ten simulation trajectories are simulated at $T_m$ and the other half are simulated at $T_m + 0.01$. The results are found to be insensitive to the small degree of superheat and the initial configuration. Furthermore, we have also performed simulations at $T_m + 0.1$ at the lowest pressure for each polygon. The patterns of the body-orientational field remain unchanged, as shown in Figure S16, while the phase transition is significantly accelerated, manifesting our conclusions are robust with respect to a wide range of temperatures. For the reverse process, the initial configurations are simulated for $1.6\times10^7$ steps at a temperature that is 0.01 lower than the corresponding phase transition point to observe the s-s transition. Once the reverse trajectory ends in polycrystalline morphology, it never equilibrates into a perfect crystal even with an extended simulation time up to $3.2\times10^7$ steps.

We use L-J units in this work, and set L-J parameters $m$ = 1, $\sigma=\sqrt[6]{2}$, and $\varepsilon=5$. The cutoff radius of the L-J interaction is set to be $r_c=7$, ensuring a high accuracy on the estimation of the potential energy. The pressure and temperature are controlled by the Nóse-Hoover barostat and thermostat[70-73], respectively. The time integration is performed by the velocity-Verlet algorithm with a timestep of 0.005. The strict 2D nature is realized by performing time integration on $x$ and $y$ directions and setting all $z$ coordinates as 0. A more concrete discussion on the selection of the parameters can be found in our previous work[48].

For the heating process, we investigated the s-s transition at various pressures to ensure our conclusions are robust. The lowest pressure we have studied is the lowest one at which the rotator crystal can stabilize, i.e., $P$ = 0 for pentagon and octagon while $P$ = 1.5 for hexagon. The highest pressure we have examined is $P$ = 10 for pentagon, $P$ = 7.5 for hexagon, and $P$ = 8 for octagon. A higher pressure for octagon can easily cause numerical instability in MD simulations as we have reported in our previous study[48], and a higher one for hexagon leads to numerical instability in fixed simulations. Only the case at $P$ = 0 has been studied for the reverse process, since a relatively high pressure may easily cause numerical instability in the simulations.

In the fixed simulation, each monomer is treated as a rigid body, whose translational motion is eliminated in each step to ensure a fixed COM. The temperature is controlled by the Langevin dynamics. The timestep in the fixed simulations is reduced to 0.001 to avoid numerical instability. Compared to the original ball-stick polygon, the motions of all the hard modes, i.e., the vibration of the bonds and the angles, are removed for a rigid body, but it can be proved that this has no influences on the thermodynamic free energy landscape[74]. Eight points were selected out from each unfixed MD trajectory to serve as the starting points to run 8 individual fixed simulations. The error-bars of the data points in Figure 2, Figure S12, and Figure S14 represent standard deviations, resulting from the thermal fluctuations in one fixed simulation trajectory.

*Data Analysis with Body-Orientation Field*: The angle of each unit vector is given by the argument of the body-orientational order parameter with respect to the $x$-axis. The *local defect* is defined by $\oint\nabla(n\theta_i)\cdot d\vec{s}=2\pi q_i$, which has the same expression as for the topological defect in 2D systems except that the integration here is taken on the smallest closed loop around each lattice point rather than globally. First provided for the studies of the XY model and 2D crystals[75], this quantity has now been applied to a variety of systems including liquid crystal[76], cell tissue[64], and amorphous solid[77,78], concentrating on diverse vector fields. As shown in Figure S3-S11, the locations of the defects highly coincide with the places where body-orientations change drastically, manifesting that the local defects capture the essential features of the amorphization process in the body-orientation field. This can also be evidenced by the synchronous time evolution of the defect concentration and the body-orientational order parameter as shown in Figure S2. Moreover, it is obvious that the positive charges are much more than the negative ones, different from the case for topological defects, where the system must stay neutral to ensure a finite energy[75], demonstrating that the local defects we define here are not topological ones.

The spatial distribution of the defects during the phase transition process is quantified by dividing the simulation box into $32\times31$ equal cells and then averaging the number of defects in all sampled configurations along the transition. We have also double-checked by dividing the box into $26\times24$ cells and found that the patterns of local defects are qualitatively the same as those shown in Figure 2. To further ensure the robustness of these patterns, we performed the same analysis for another three sets of individual trajectories under the same conditions at the lowest pressures, and found no qualitative differences.

*Finite Size Effects*: All the results reported in the main text were obtained from the simulations with 2496 ($52\times48$) monomers. We further performed larger simulations with 4620 ($70\times66$) monomers to estimate the finite-size effect. As shown in Figure S17 in SI, the pattern for each polygon is qualitatively the same. Specifically, the striped region for pentagon is clearer in the larger system, which further verifies the vague striped region in the body-orientation field shown in Figure 1(d). Moreover, there can be two striped regions in the larger pentagon system, since a larger system can provide more 'nucleation sites'. More discussions can be found in SI.

**Acknowledgements**

The computations of this work were conducted on the HPC cluster of ITP-CAS, Sugon-Taiyuan, Sugon-Xi'an, Sugon-Wuzhen, and Tianhe-2 supercomputer. This work was supported by the National Natural Science Foundation of China (No.12047503).

**Author Contributions**

R. Z. conceived and designed the research. R. Z. performed the simulation and analyzed the data under the supervision of Y. W. All the authors discussed the results and wrote the paper.

**Data Availability Statement**

The data that support the findings of this study are available from the corresponding author upon request.

**Supporting Information**

Supporting Information is available from the Wiley Online Library or from the author.

# Supplementary Information for: Shape-Determined Kinetic Pathways in 2D Solid-Solid Phase Transitions

Ruijian Zhu[1,3], Yi Peng[2,3], Yanting Wang[1,3] .

[1] Institute of Theoretical Physics, Chinese Academy of Sciences, Beijing 100190, China
[2] Beijing National Laboratory for Condensed Matter Physics, Institute of Physics, Chinese Academy of Sciences, Beijing 100190, China
[3] School of Physical Sciences, University of Chinese Academy of Sciences, 19A Yuquan Road, Beijing 100049, China

**This PDF file includes:**

Supplementary Text
Figure. S1 to S17
Table S1
Reference 1

## S1. Robustness of Body-Orientation Fields at Various Pressures

In Table S1, we list the differences of thermal quantities under various pressures, manifesting that the structures of two solid states are more similar at a higher pressure. This demonstrates that the energy barrier for rotational motion decreases as pressure increases. Moreover, the body-orientational order parameter for pentagon takes a value about 0.7 in the parent phase, much less than about 0.85 for hexagon and octagon, indicating that local defects are easier to be created even before transition, as evidenced in Figure S3(a). It is worth mentioning that the s-s transition of pentagon and octagon can occur at an arbitrary pressure including $P$ = 0, but the one of hexagon occurs only at $P \geq 1.5$. This is because the close-packing phase of hexagon is ultra-stable at low pressures, as a consequence of the perfect match of interaction and shape, allowing it stabilizes up to a relatively high temperature, and then melts into fluid state directly[1].

In Figure S3-S11, we show the time evolution of the body-orientation field for each polygon at various pressures. In each figure, the pointing direction of each arrow represents the argument of the body-orientation order parameter $\phi$ with respect to the $x$-axis, while the purple and blue dots label the positions of local defects with positive and negative charges, respectively. It is evident that the majority of the dots in these figures are colored in purple, suggesting that the system has a net positive 'charge'. This implies that the local defects we define here are not topological defects, but rather indicators of the regions where body-orientations change dramatically with respect to neighboring monomers.

The time evolution of the local defects described in the main text is robust at different pressures, as demonstrated by performing statistical analysis on the spatial distribution of local defects. The results are depicted in Figure S12, where pentagon exhibits one or two vague stripe regions with several isolated warm-colored points, hexagon shows no collective patterns, and octagon consistently displays a distinct stripe region. The fact that pentagon can have more than one stripe region should be attributed to the lower energy barrier at higher pressures, which also slightly fuzzes up the stripes.

In Figure S3-S11, we also label the corresponding MD steps for each panel. Comparing the time intervals between different panels in Figure S5, Figure S8, and Figure S11, which present the evolution of the body-orientation fields for octagon at different pressures, one can identify that the time required for forming a narrow stripe region across the box is very short and quite stable at different pressures, but further growth along the normal direction is significantly accelerated at higher pressures, consistent with the fact that the energy barrier for rotational motion is lower at higher pressures. Recalling that the thermodynamic quantities of pentagon have the same tendency as those of octagon with increasing pressure, it is expected that the growth rate for pentagon should follow the same trend as octagon. This can indeed be observed as in Figure S3, S6, and S9, despite the smearing of the stripe regions. As shown in Figure S4, S7, and S10, the behavior of hexagon is entirely different: The transition process is dominated by translational motion and no stripe regions appear due to its weak dependence on the energy barrier, leading to an almost constant transition rate invariant to pressure.

## S2. Details of Fixed Simulations and Robustness of Kinetic Pathways

In the fixed simulations for each polygon at each pressure, the initial configurations are selected along the original unfixed MD trajectory with the occurrence of the s-s transition.

This is illustrated at the lowest pressures for different polygons in Figure S13(a)-(c), in which each initial configuration for the corresponding fixed simulation is marked as a colored point along the time-evolution curve for the body-orientational order parameter in the unfixed (original) MD trajectory. The time evolutions of the body-orientational order parameter in the fixed simulations at the lowest pressures for different polygons are shown in Figure S13(d)-(f). The same procedure is employed for the fixed simulations at other pressures, with the time evolutions of the body-orientational order parameter being qualitatively the same. In Figure S14, we plot the results from the fixed simulations at various pressures for each polygon, demonstrating the robustness of the kinetic pathways with respect to thermal conditions. There are fewer available data points at higher pressures because numerical instability causes the fixed simulations to have a strong tendency of breaking down. The fixed simulations for the reverse processes follow the same procedure, whose results are shown in Figure S15. It can be seen that a polygon system may have more than one kinetic pathway for the reverse process. However, it is evident that a kinetic pathway dominated by translational motion at the initial stage consistently traps into a polycrystalline morphology, and always ends up with a perfect crystal state otherwise. All the average values for the fixed simulations shown here and in the main text are calculated after the constrained systems are equilibrated.

## S3. Finite-Size Effects

To eliminate possible finite-size effects, we further performed simulations on the systems composed of 4620 polygons at $P = 0$ for pentagon and octagon as well as at $P = 1.5$ for hexagon. These simulations follow the same procedure as the ones for 2496 polygons. As shown in Figure S17, the patterns of local defects in the body-orientation field for different polygons are qualitatively the same. Specifically, pentagon can exhibit one or two crossed striped regions, as depicted in Figure S17(a) and (b), since a larger system provides more ‘nucleation sites’, enabling the spontaneous formation of two striped regions, benefiting from the relatively easy rotational motion. Furthermore, the striped regions in these two panels are clearer than the ones in Figure 2(a), Figure S12(a), and Figure S12(d), manifesting that the formation of striped region persists in larger systems.

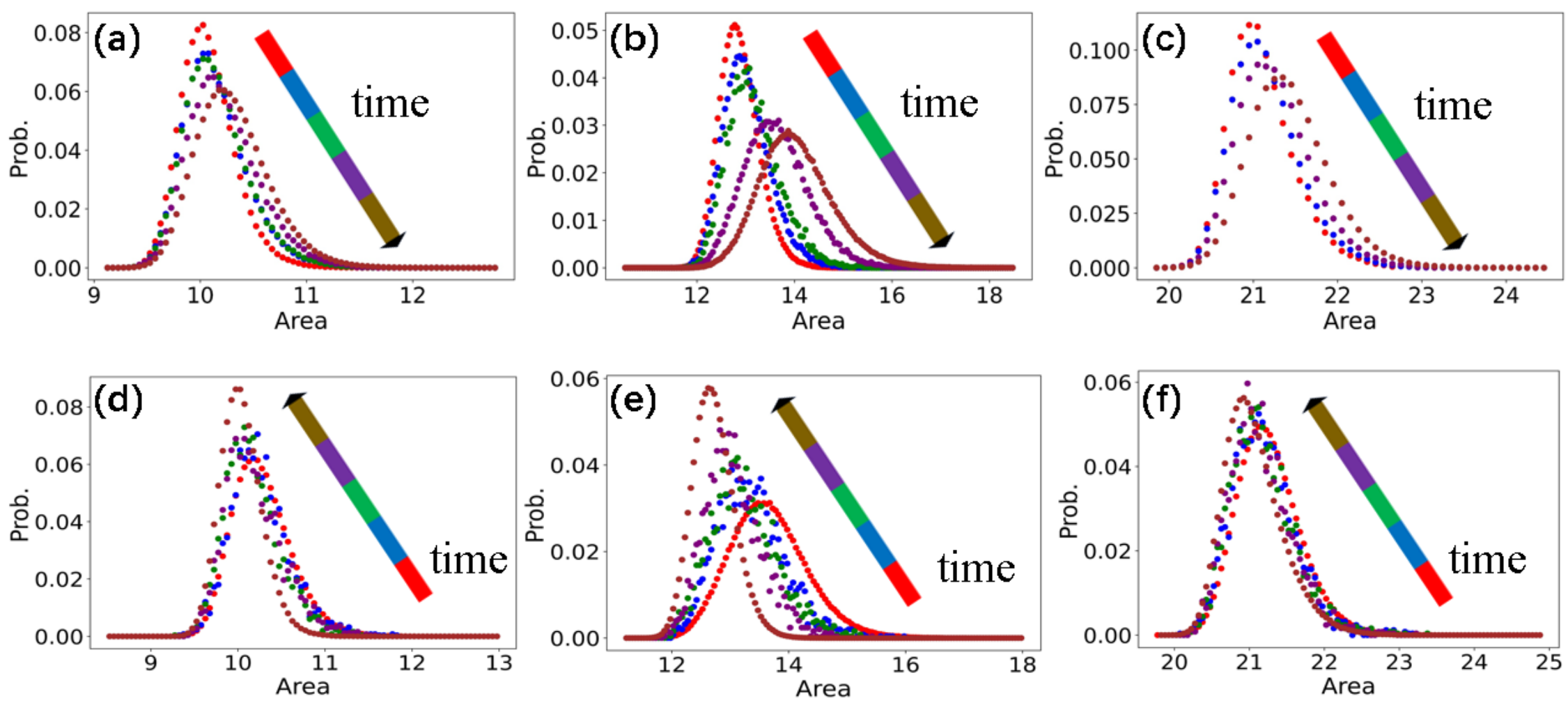

**Figure S1. Time evolutions of the distributions of Voronoi cell area in the heating (a)-(c) and cooling (d)-(f) processes of s-s transition for pentagon, hexagon, and octagon.** (a) and (d) for pentagon, (b) and (e) for hexagon, (c) and (f) for octagon. The probability density (Prob.) always exhibits a single-peak distribution, while the peak value moves continuously towards the direction corresponding to a larger (for heating process) /smaller (for cooling process) cell area, which is in consistent with the theoretical prediction of homogeneous expansion/contraction for isostructural s-s transition.

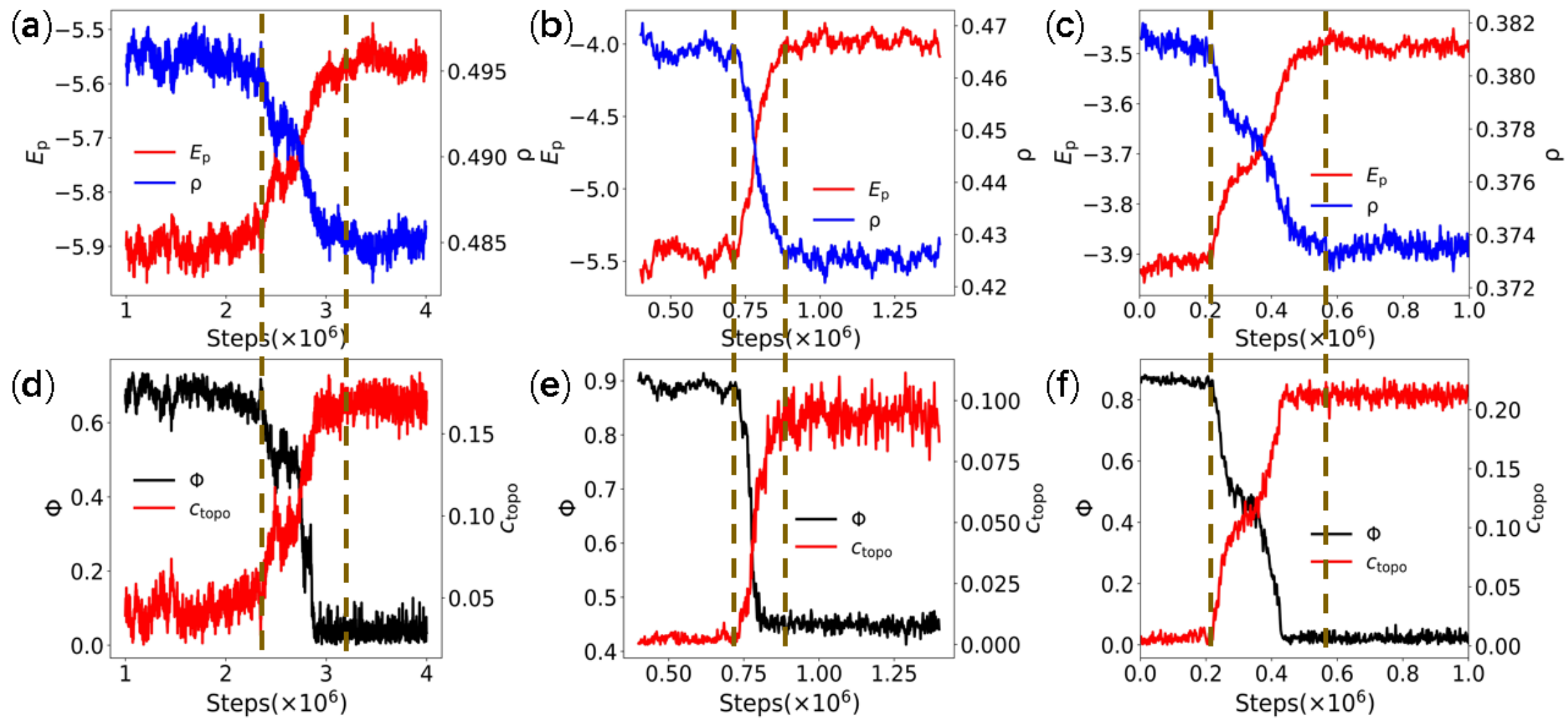

**Figure S2. Time evolutions of the collective quantities during s-s transition.** (a)-(c) for the evolution of potential energy and density, (d)-(f) for the evolution of body-orientational order parameter and concentration of local defects. (a) and (d) for pentagon, (b) and (e) for hexagon, (c) and (f) for octagon. The dashed brown lines are used to guide the eyes for the initiating and ending points of the transition. It can be seen that different quantities start to rise up at almost the same time, while the potential energy and density take several more steps to approach their final values than the body-orientational order parameter and concentration of local defects. It is also evidenced that the concentration of local defects varies synchronously with the body-orientational order parameter.

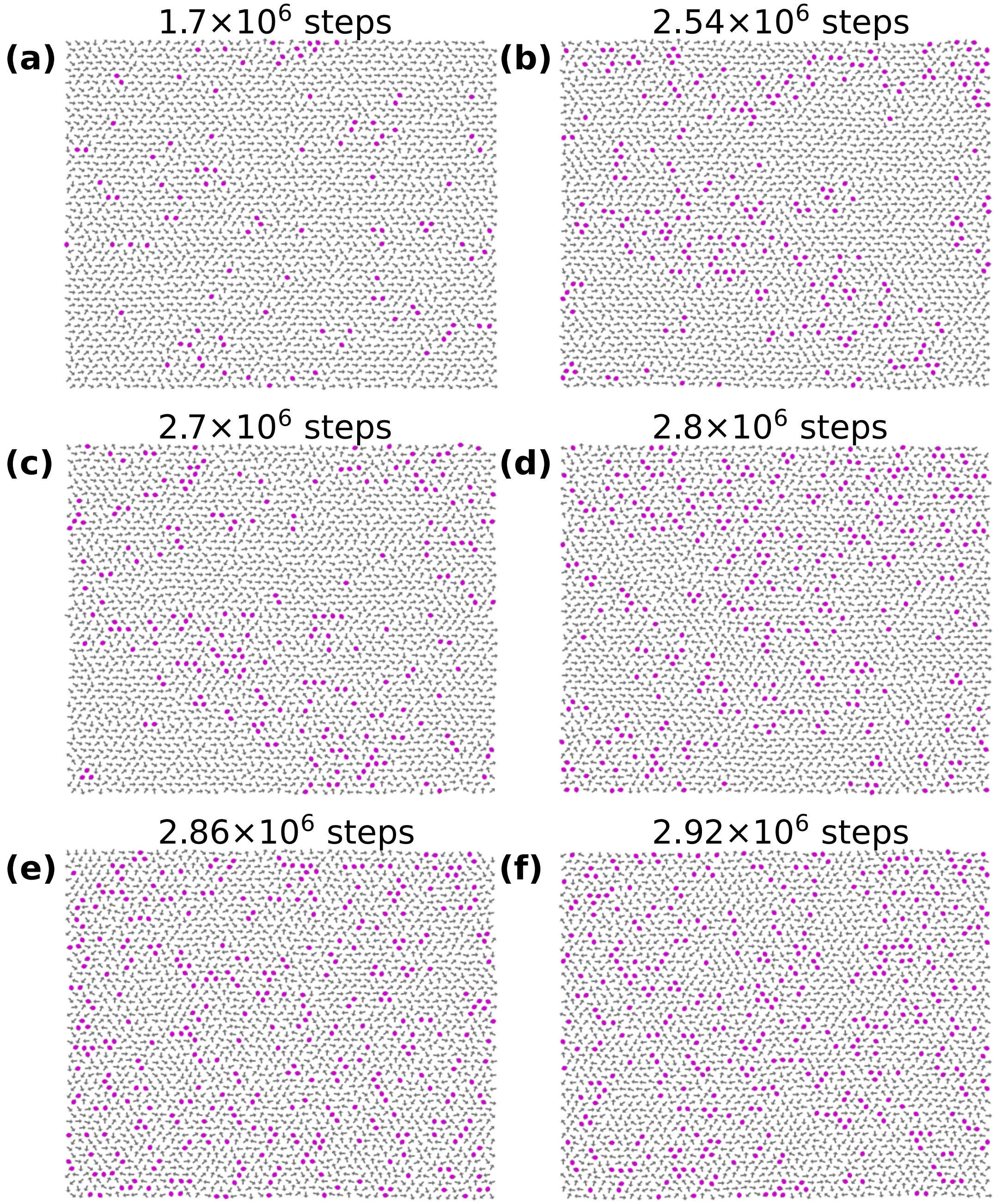

**Figure S3. Defects of pentagon at *P* = 0.** Each arrow located on the COM of a ball-stick polygon represents the argument of the corresponding body-orientational order parameter with respect to the *x*-axis. The dots mark the positions of the local defects in the body-orientational field, colored in purple and blue (can hardly be seen here) dots for positive and negative 'charges', respectively. The same plotting scheme is also applied to Fig. S2-S9. The parent phase shown in (a) already contains several local defects. After forming a vague stripe region shown in (b), the number of defects increases with time, as shown in (c)-(e). Finally, as shown in (f), the system transforms into a rotator crystal with a large number of uniformly distributed local defects.

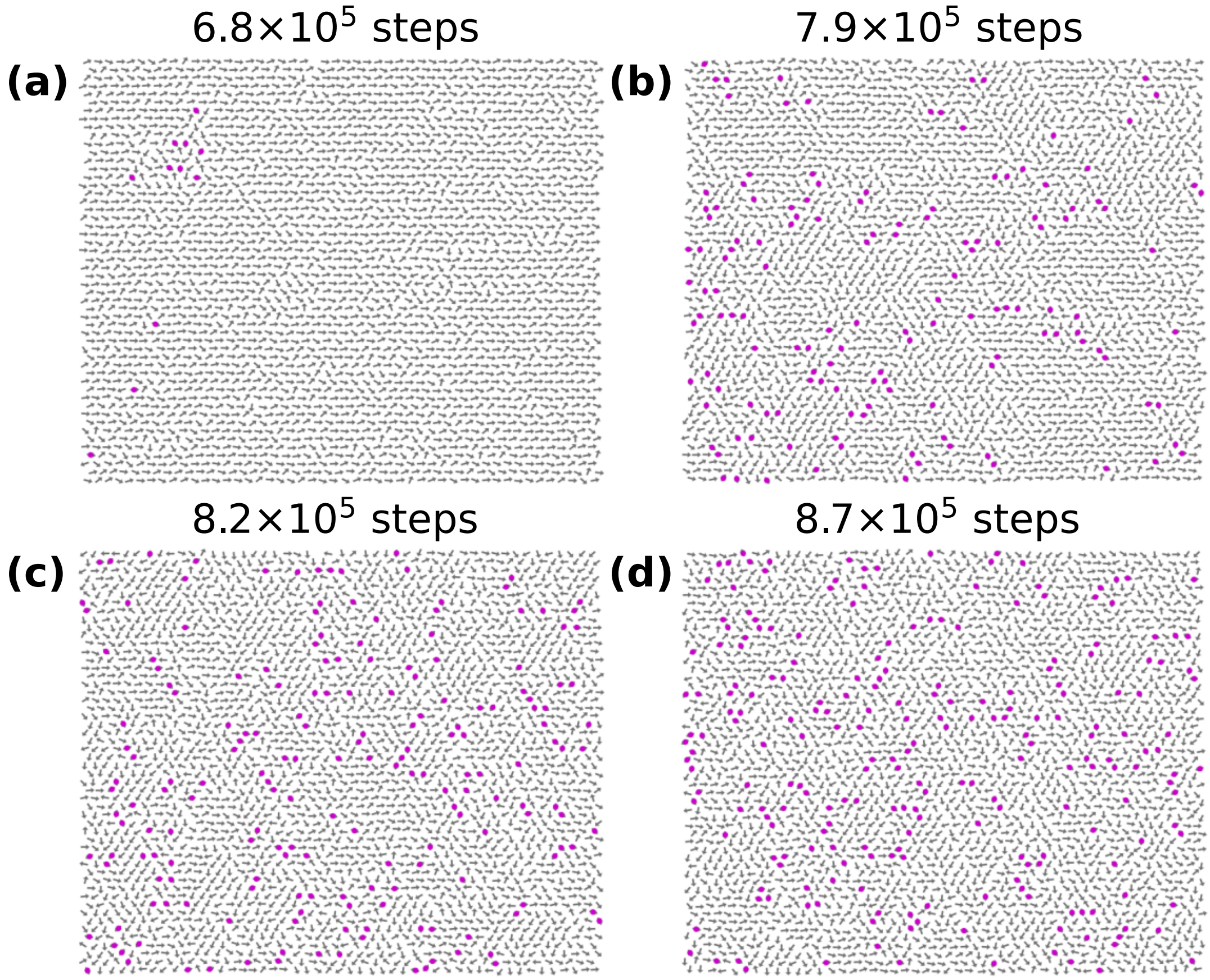

**Figure S4. Defects of hexagon at *P* = 1.5.** There are very few defects at the initial stage shown in (a). The number of defects then increases with time during the transition process, as shown in (b)-(d), with no evidence of a collective pattern.

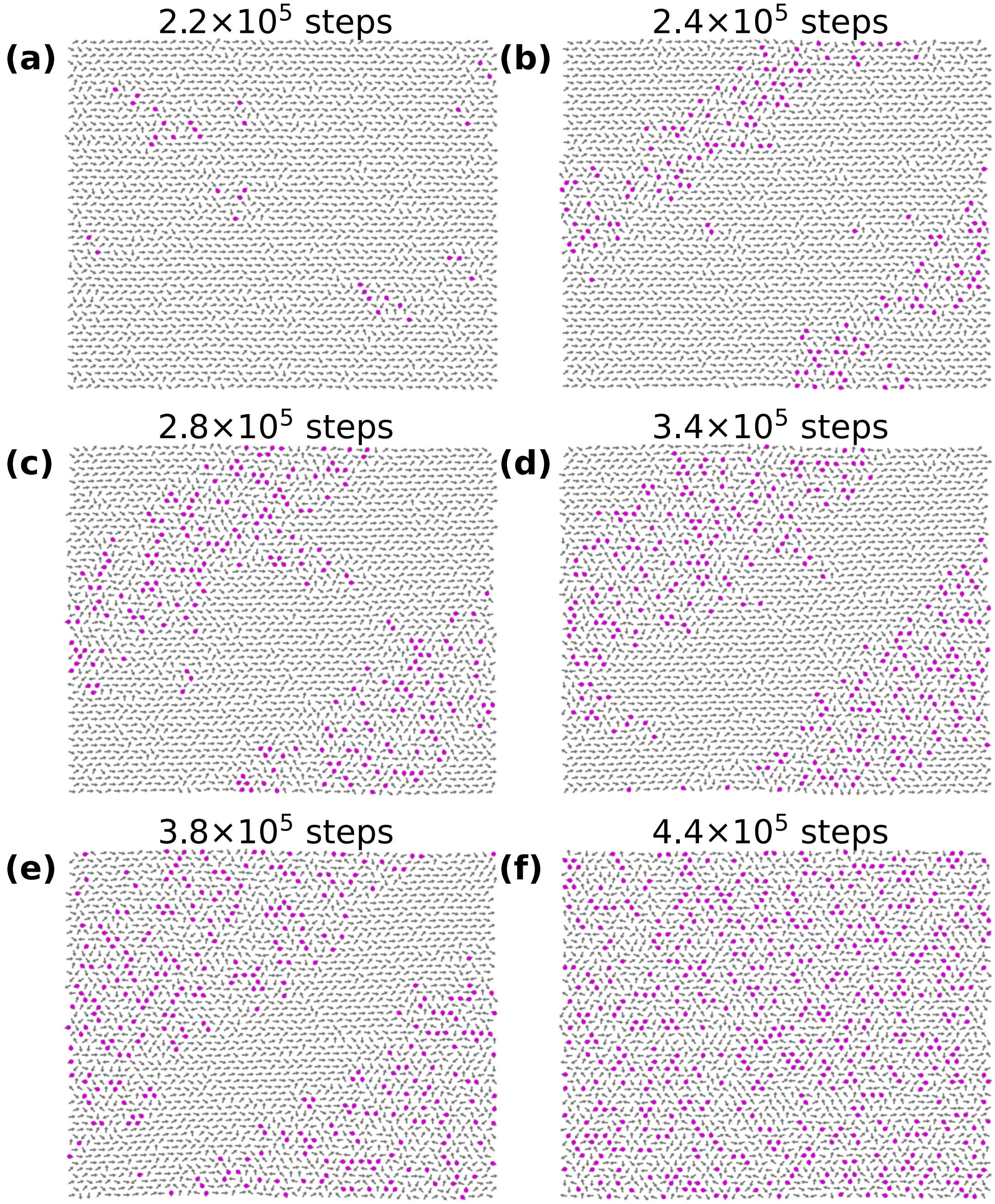

**Figure S5. Defects of octagon at *P* = 0.** Similar to hexagon, only a few defects present at the initial stage, as shown in (a). As shown in (b), after a very short time, the defects form a narrow stripe region across the periodic simulation box. Then the stripe region attempts to grow along its normal direction, as shown in (c)-(e). Finally, as shown in (f), when the stripe uniformly covers the whole simulation box, the system transforms into the rotator crystal phase.

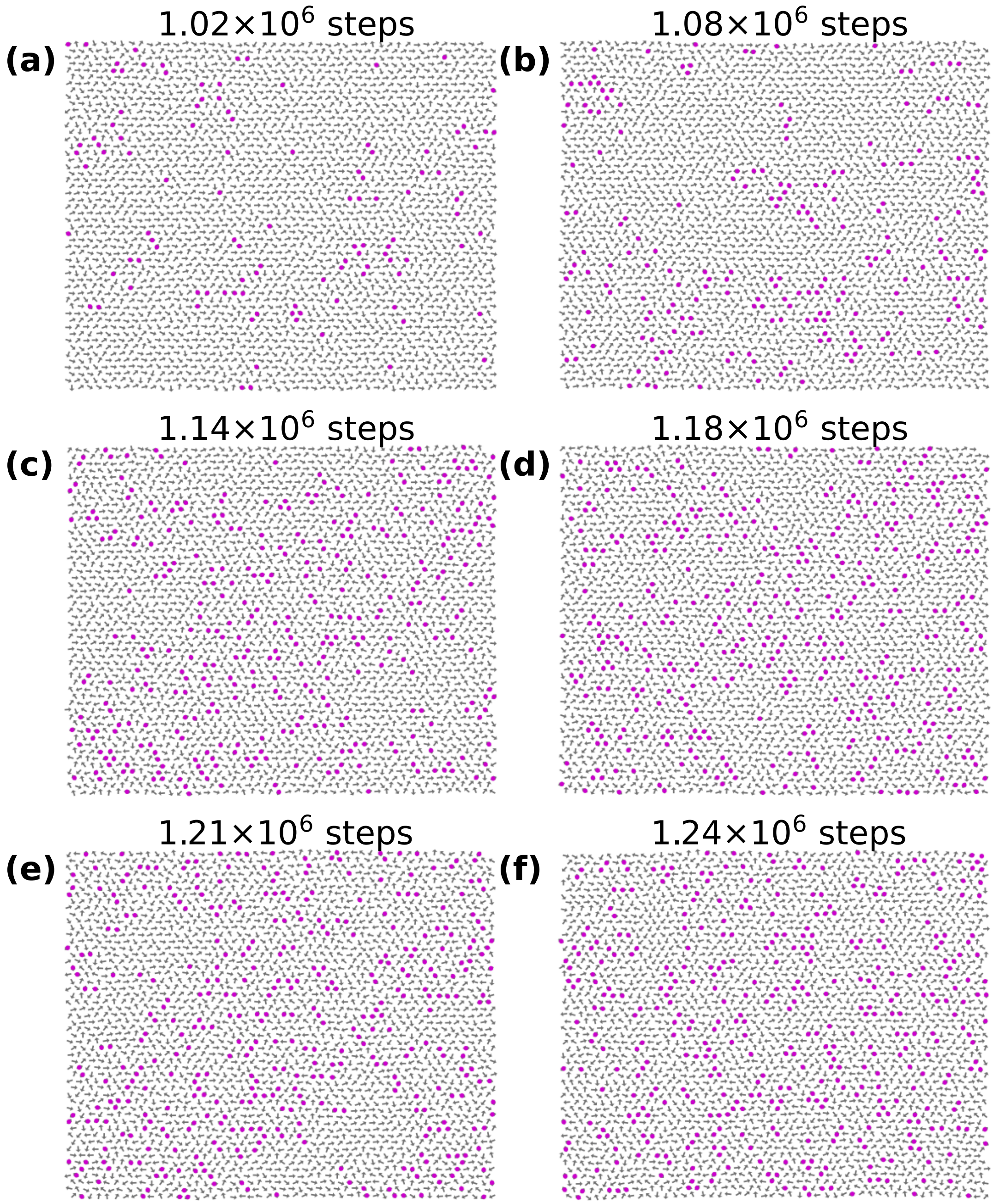

**Figure S6. Defects of pentagon at *P* = 5.** It is similar to the case at $P = 0$, but the growth after forming a vague stripe region is significantly accelerated.

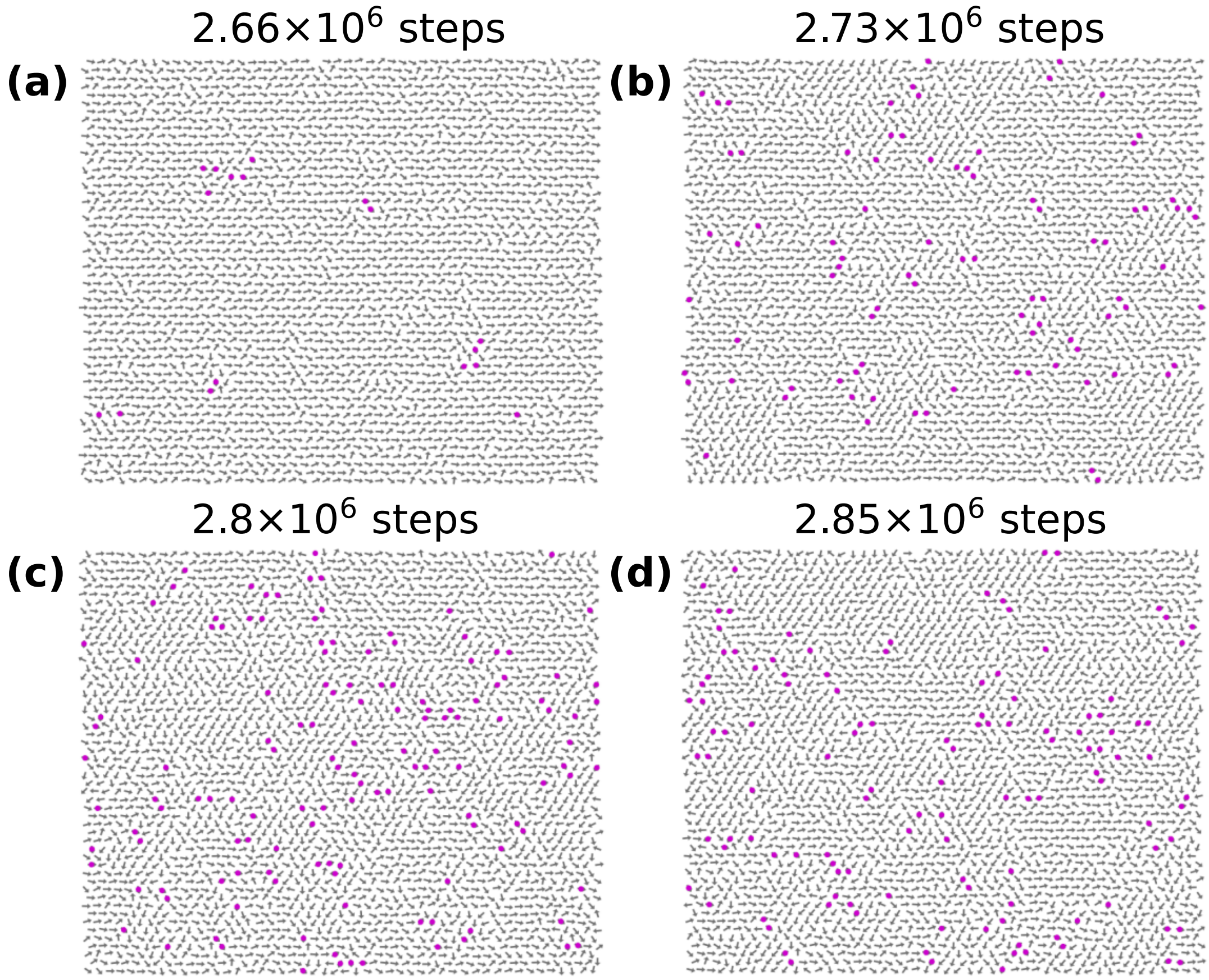

**Figure S7. Defects of hexagon at *P* = 5.** It is similar to the case at *P* = 0.

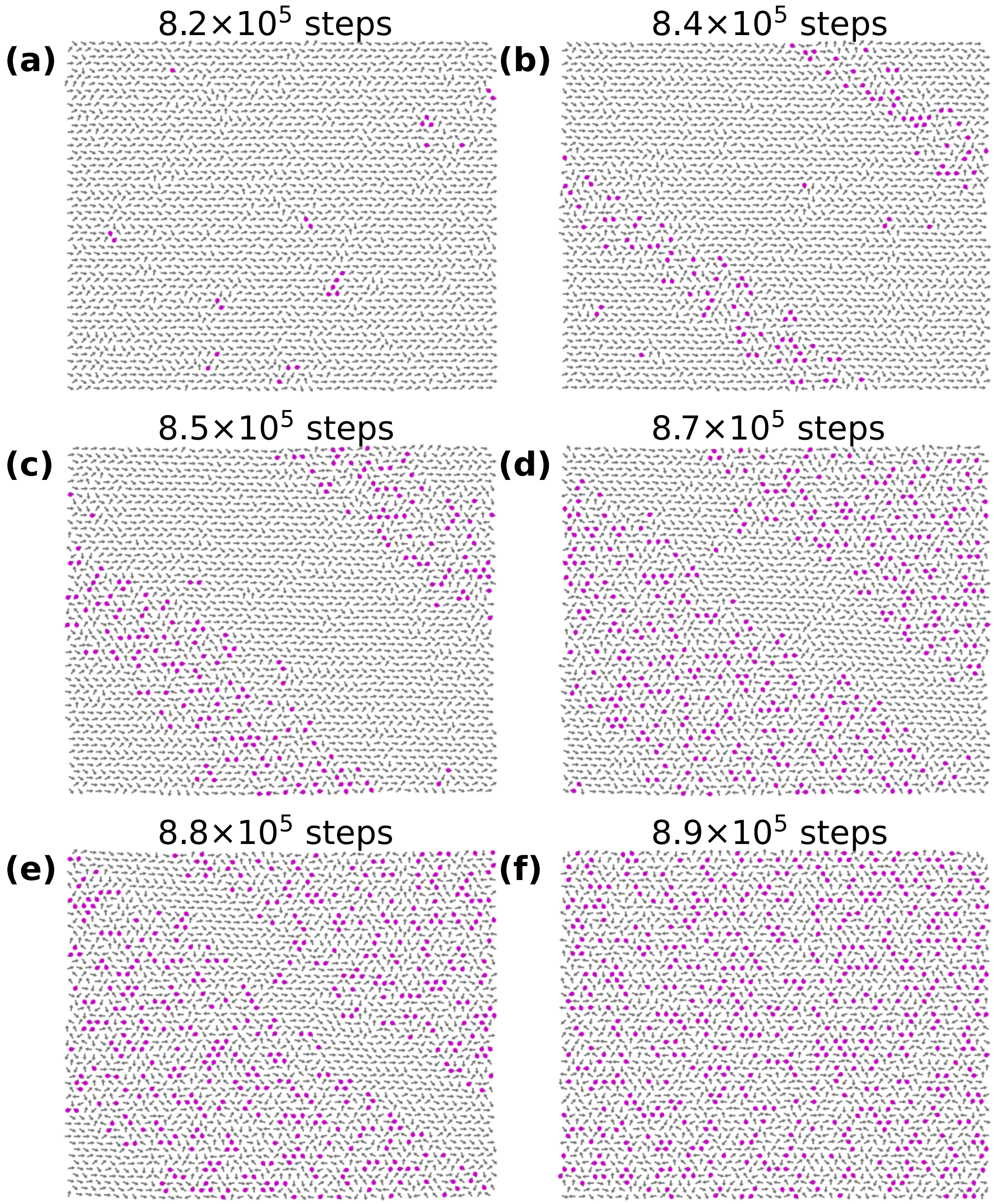

**Figure S8. Defects of octagon at *P* = 5.** It is similar to the case at *P* = 0, but the growth of the stripe region along its normal direction is significantly accelerated.

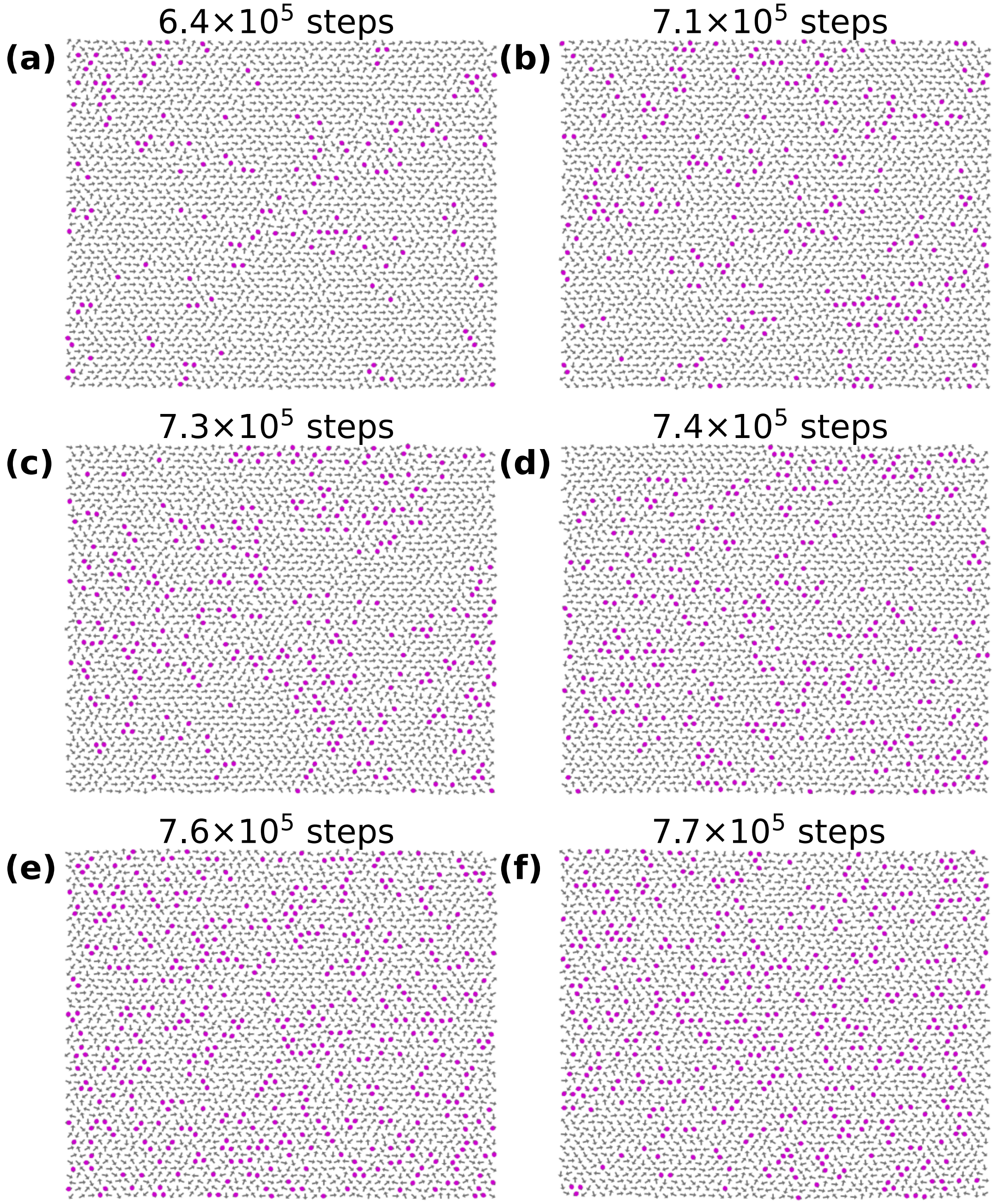

**Figure S9. Defects of pentagon at *P* = 10.** It is similar to the case at lower pressures, but the growth after forming a vague stripe region is further accelerated.

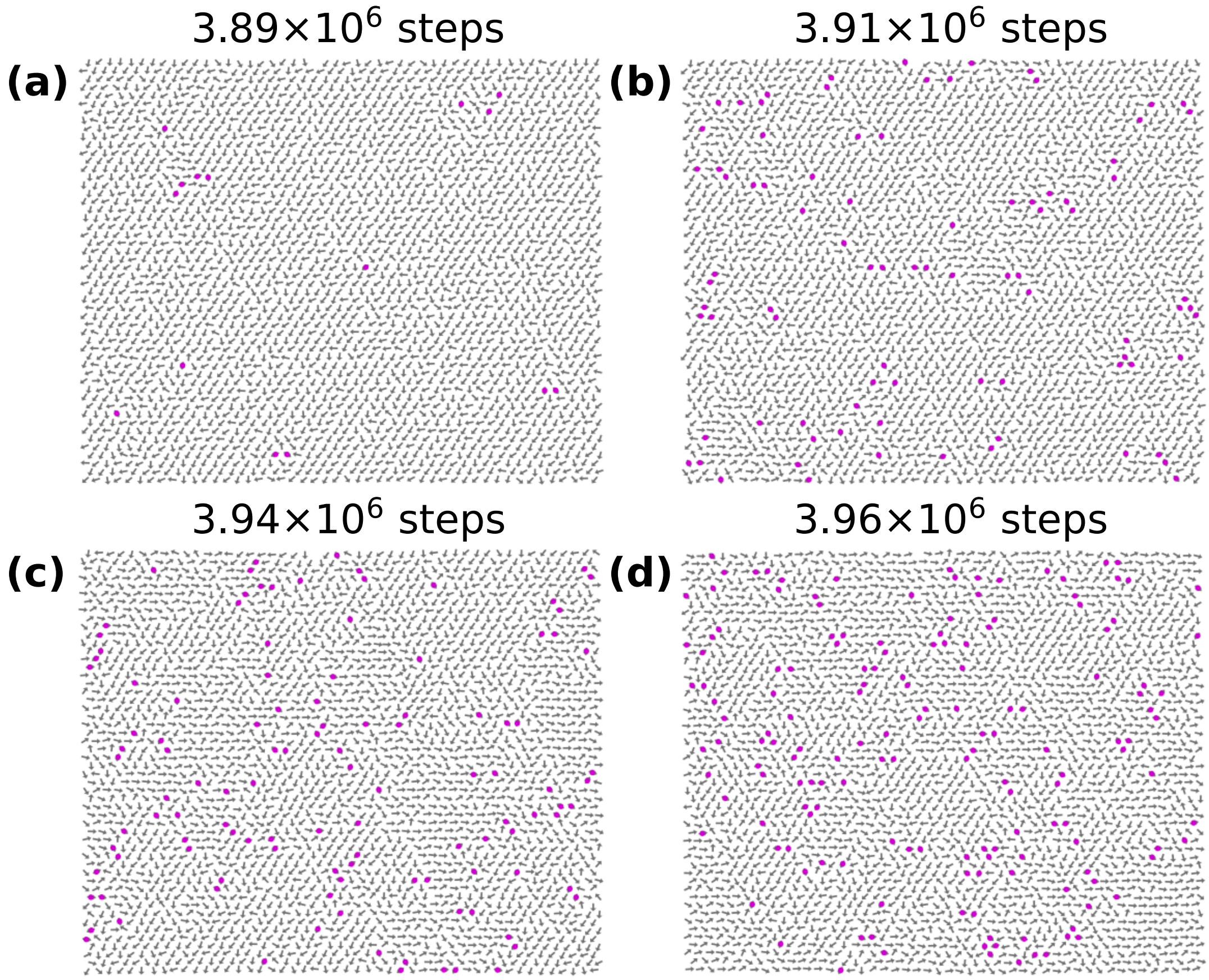

**Figure S10. Defects of hexagon at *P* = 7.5.** It is similar to the case at lower pressures.

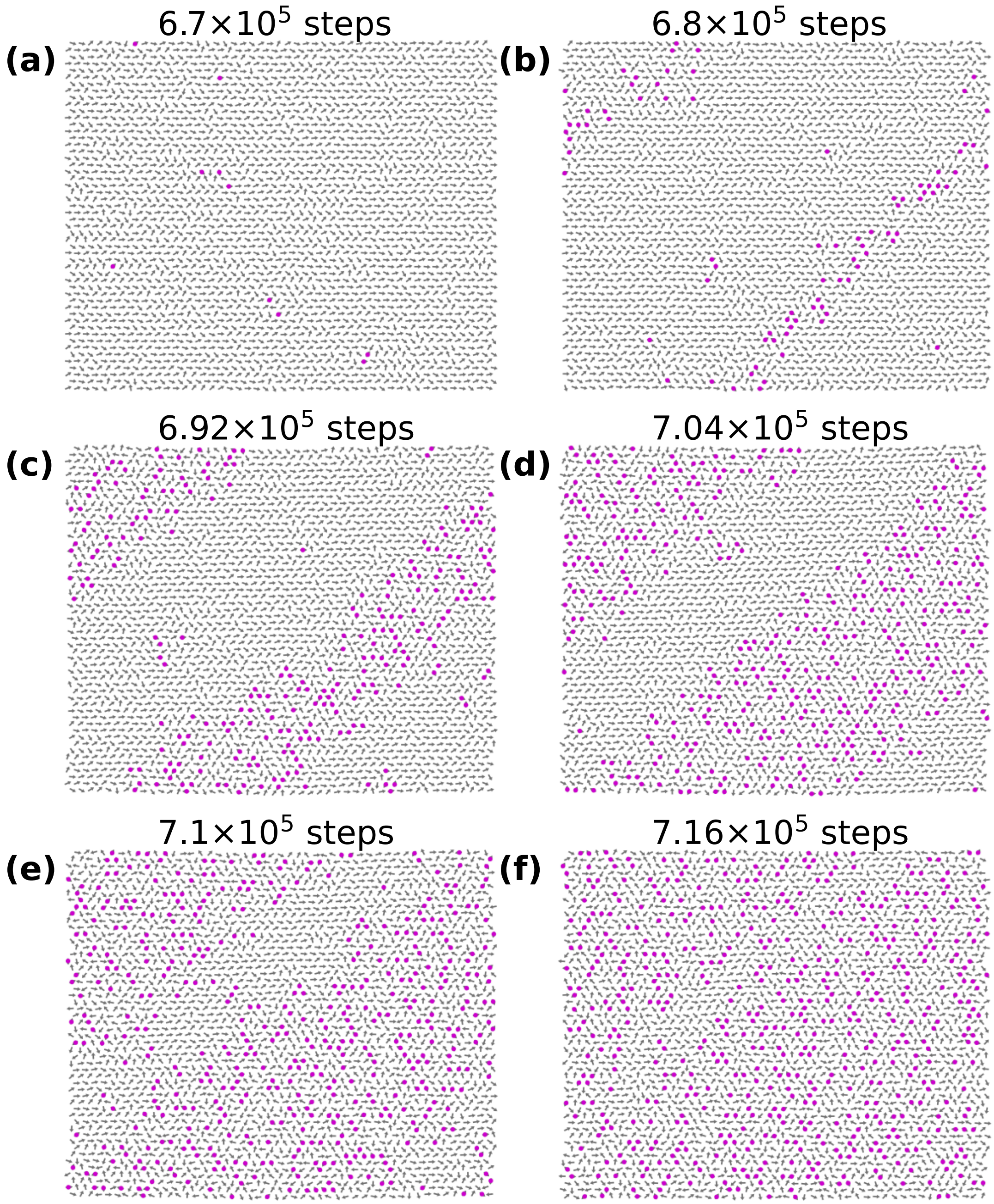

**Figure S11. Defects of octagon at *P* = 8.** It is similar to the case at lower pressures, but the growth after forming a vague stripe region is further accelerated.

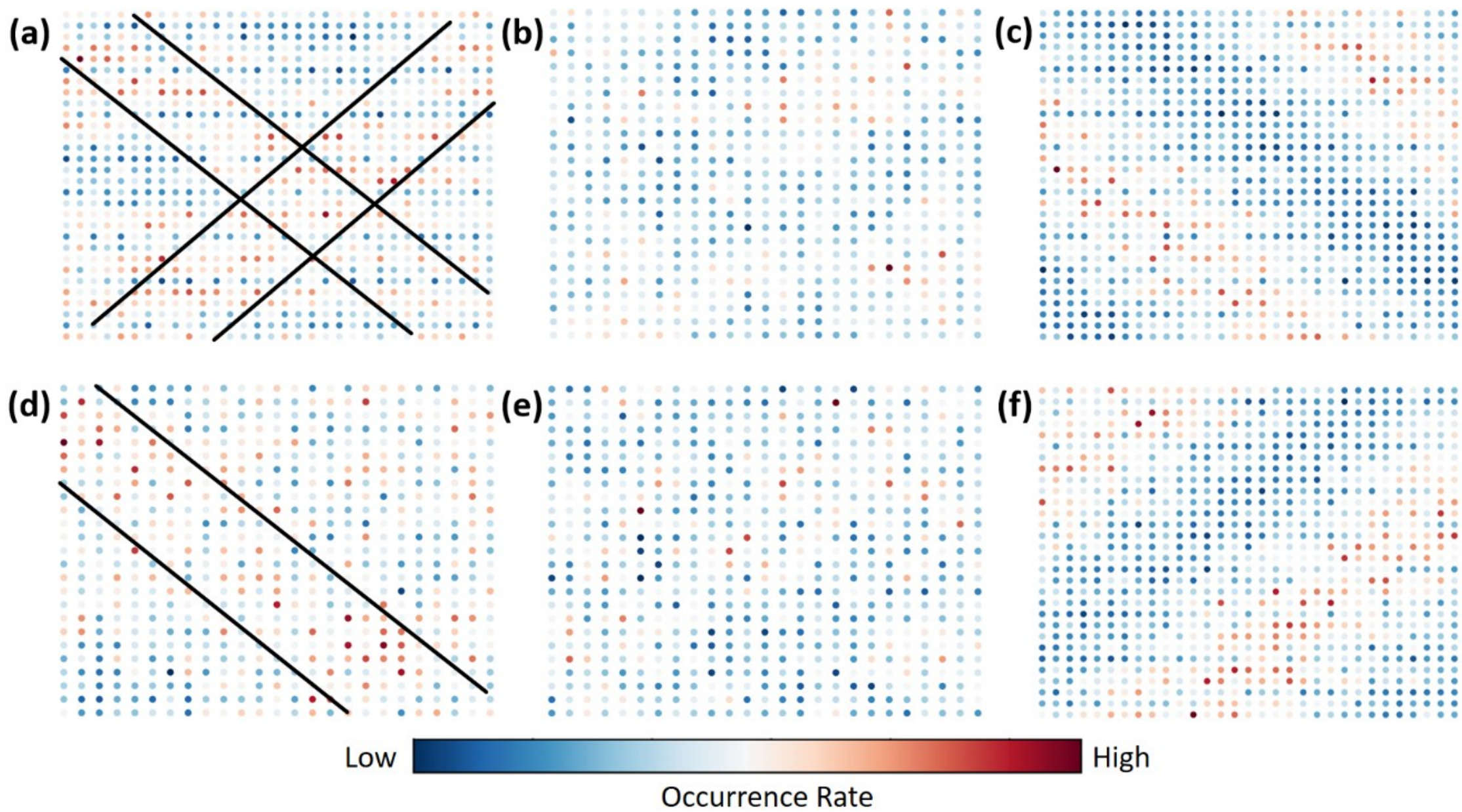

**Figure S12. Statistics of local defects at various pressures.** (a) Pentagon at $P$ = 5. (b) Hexagon at $P$ = 5. (c) Octagon at $P$ = 5. (d) Pentagon at $P$ = 10. (e) Hexagon at $P$ = 7.5. (f) Octagon at $P$ = 8. In (a) and (d), the black lines are drawn to highlight the vague stripe regions.

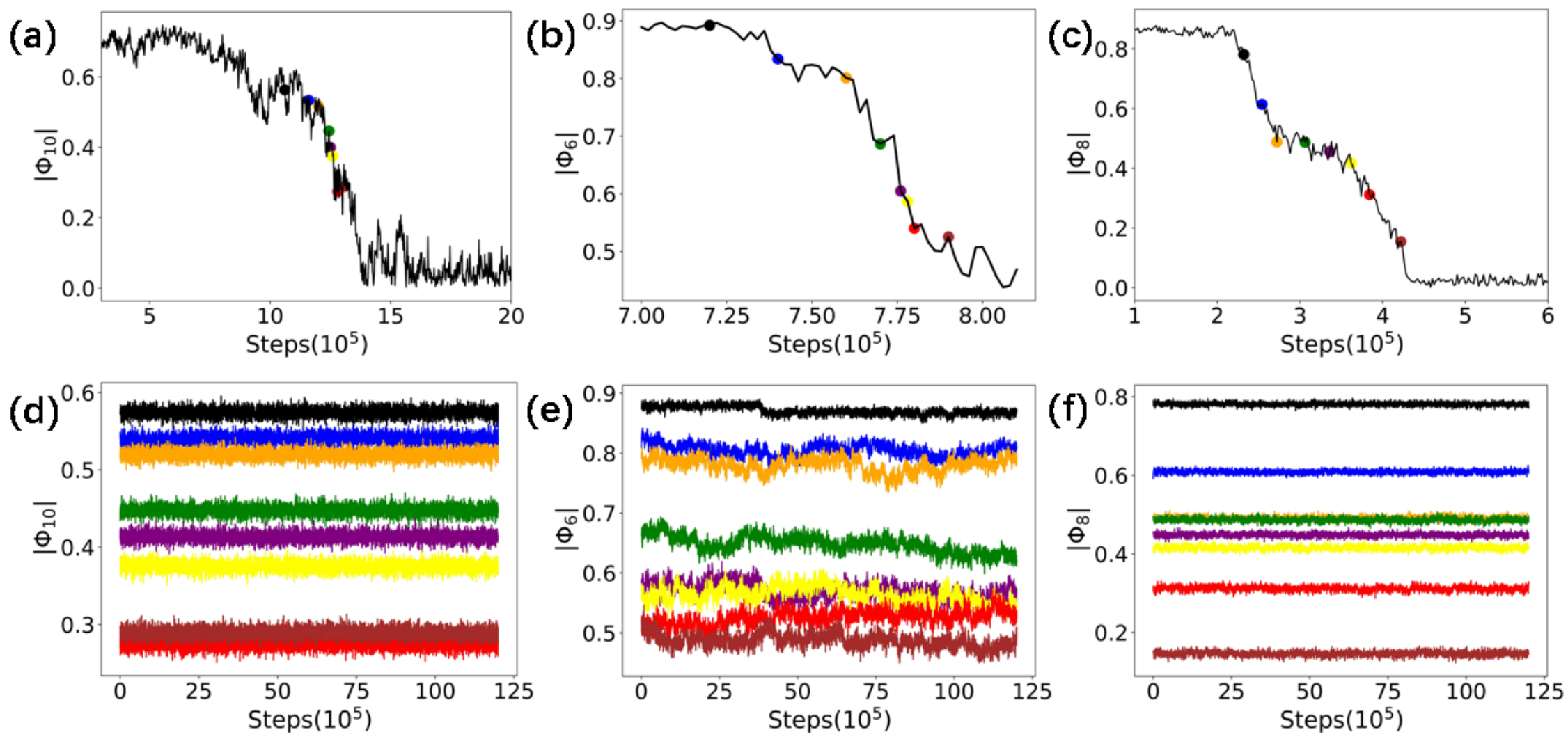

**Figure S13. Selection of initial configurations and simulation results of the fixed simulations at the lowest pressures.** In (a) - (c), the black line is the time evolution of the body-orientational order parameter, and the colored points mark the moments when the initial configurations for the fixed simulations are selected. In (d) - (f), we show the time evolutions of the body-orientational order parameter in the 'fixed' simulations, (a) and (d) for pentagon, (b) and (e) for hexagon, and (c) and (f) for octagon. The coloring of the lines in (d) - (f) is the same as the points in (a) - (c).

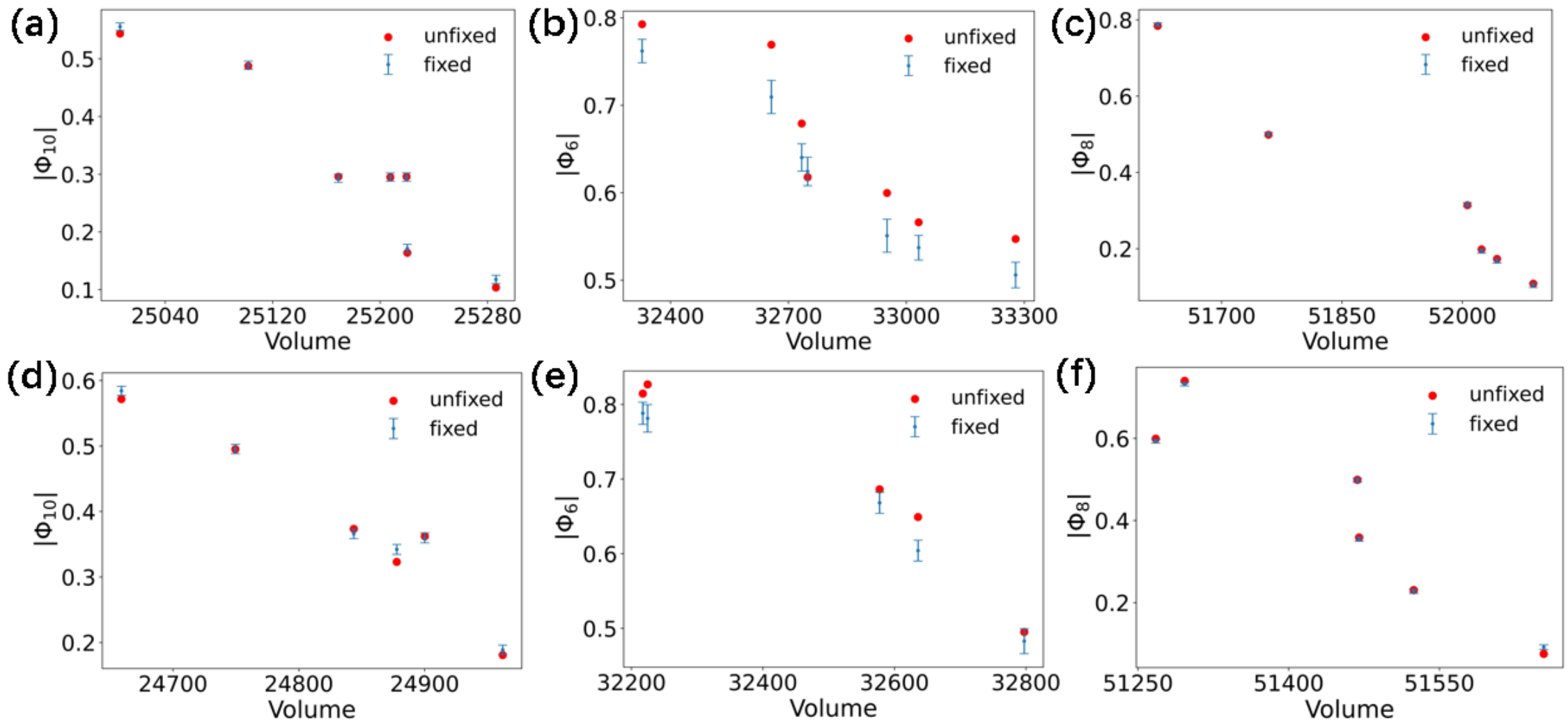

**Figure S14. Results from the fixed simulations at various pressures for the heating process.** (a) Pentagon a $P$ = 5. (b) Hexagon at $P$ = 5. (c) Octagon at $P$ = 5. (d) Pentagon at $P$ = 10. (e) Hexagon at $P$ = 7.5. (f) Octagon at $P$ = 8. Compared with the ones shown in Figure 2 in the main text, it can be found that the kinetic pathway of a specific polygon is independent of thermal conditions.

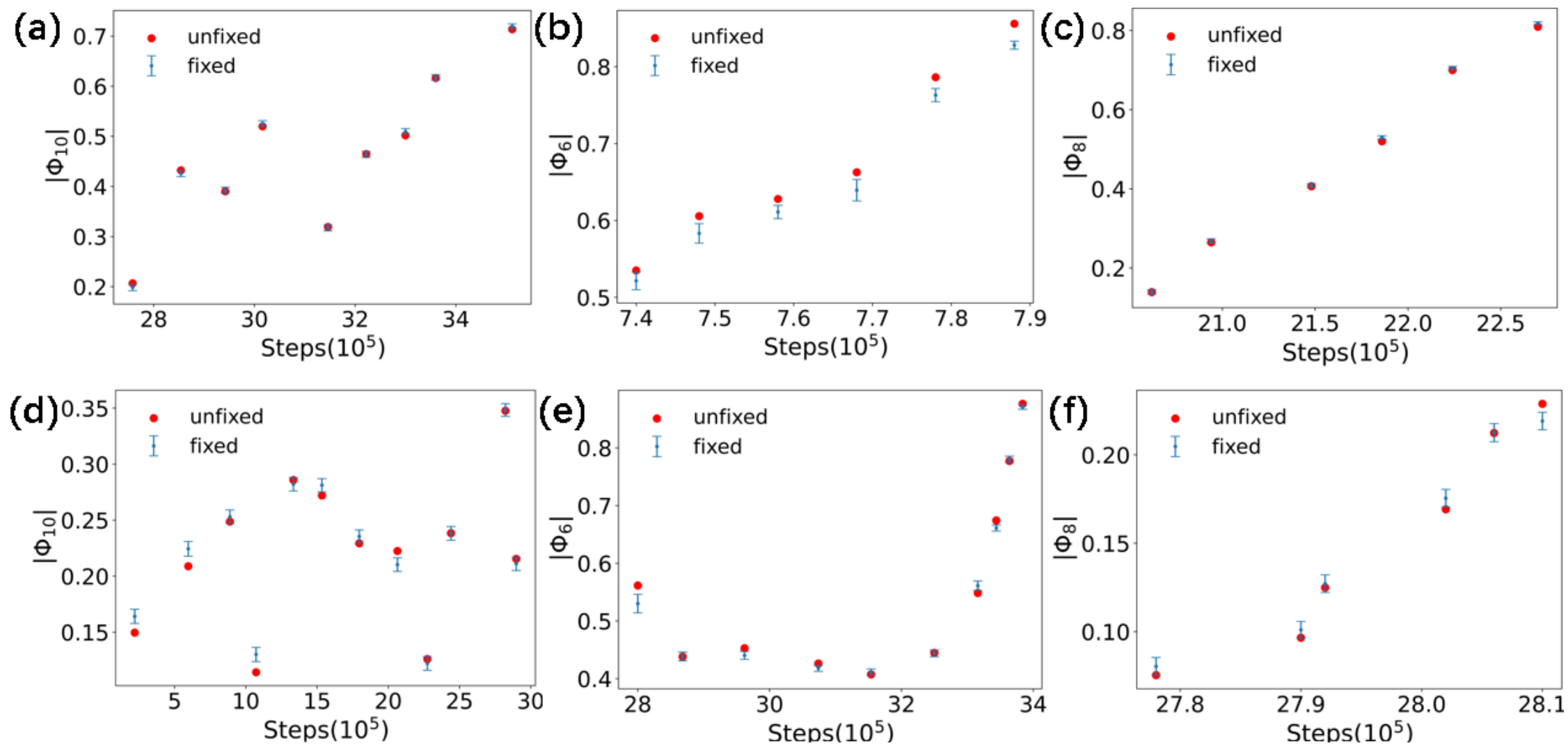

**Figure S15. Body-orientational order parameters in the fixed simulations for the reverse process.** The 'steps' are those in the unfixed MD simulation. Since some of the trajectories are trapped into the polycrystalline state, only the comparison at the early stage is meaningful, which determines whether the system can modify its body-orientational order before contracting into a dense state. (a) and (d) for pentagon, (b) and (e) for hexagon, (c) and (f) for octagon.

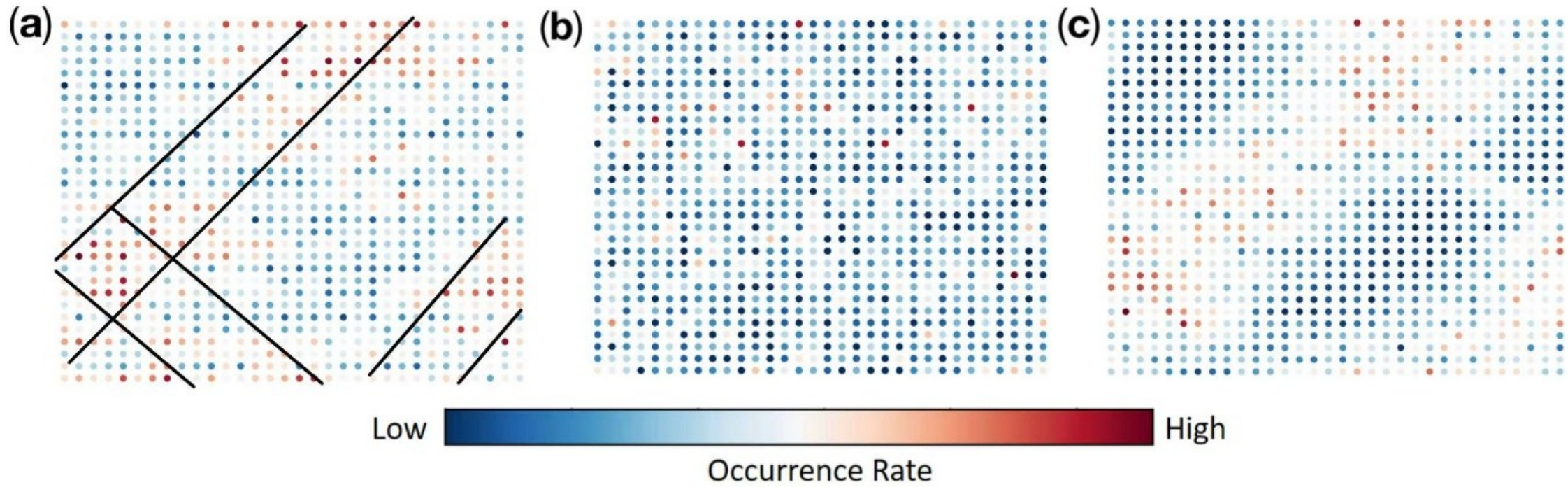

**Figure S16. Statistics of local defects in the body-orientation fields at the temperature of $T_m$ + 0.1.** (a) Pentagon, (b) Hexagon and (c) Octagon. The black lines are used to guide the eyes for the stripes. It can be found that the patterns are more smeared but qualitatively the same as the cases at a lower degree of superheat.

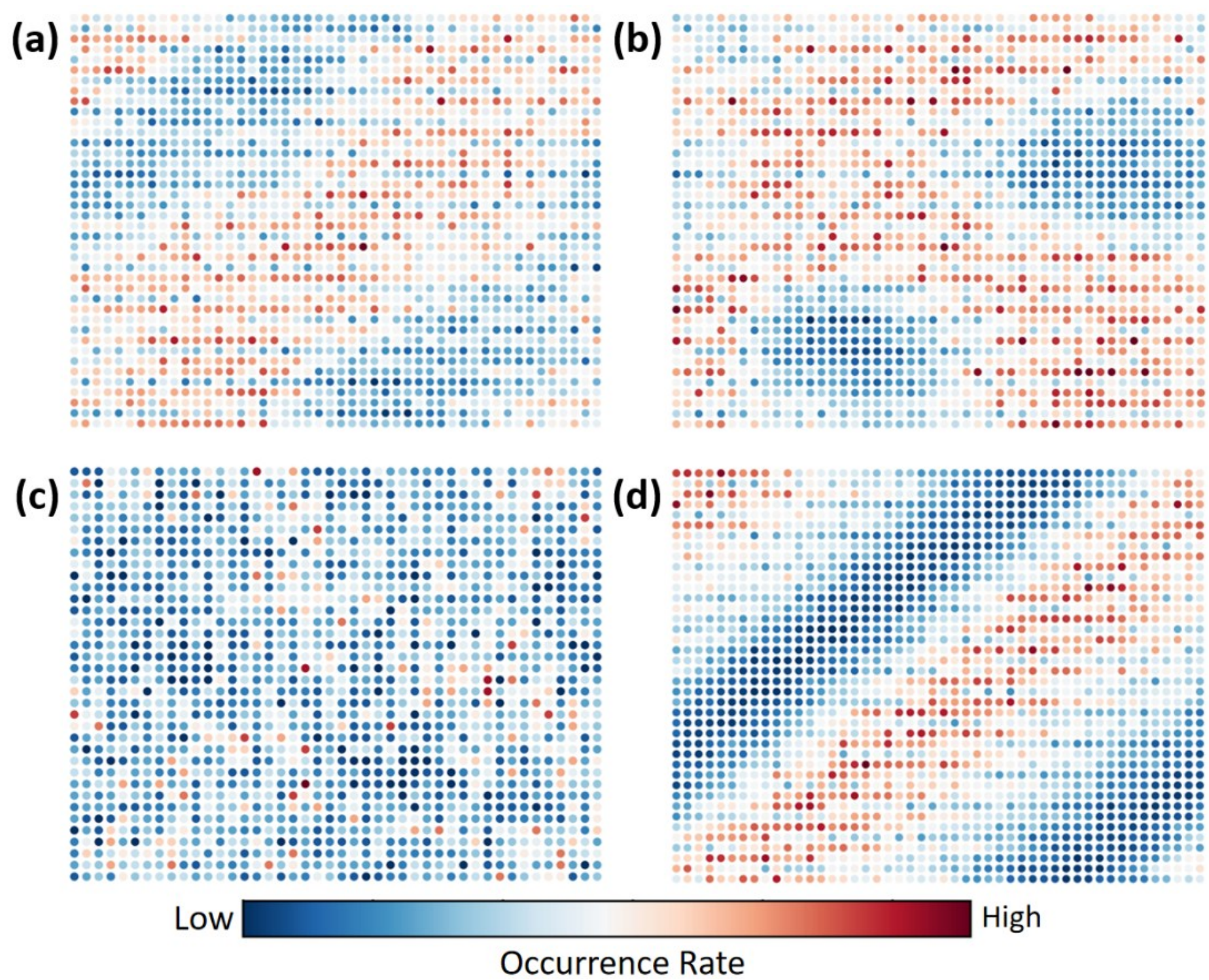

**Figure S17. Statistics of local defects in the body-orientation fields for *N* = 4620**. (a) and (b) are for pentagon, indicating that it is possible to form one or two stripe regions in a larger system. (c) for hexagon and (d) for octagon.

**Table S1. Differences of thermodynamic quantities between the parent phase and the product phase at the melting points under various pressures.**

| Shape | Pressure | $\Delta E_{\mathrm{p}}$ | $\Delta\rho^{-1}$ | $\Delta\Phi$ |
|---|---|---|---|---|
| **Pentagon** | $P = 0$ | 0.38531 | 0.049953299 | 0.71041631 |
| | $P = 5$ | 0.31036 | 0.038348263 | 0.66040313 |
| | $P = 10$ | 0.29444 | 0.035007953 | 0.65213229 |
| **Hexagon** | $P = 1.5$ | 1.449 | 0.195133792 | 0.443682344 |
| | $P = 5$ | 0.77023 | 0.090753697 | 0.415460196 |
| | $P = 7.5$ | 0.57863 | 0.066392551 | 0.395135211 |
| **Octagon** | $P = 0$ | 0.42008 | 0.052412336 | 0.851954527 |
| | $P = 5$ | 0.35332 | 0.032927826 | 0.8297257 |
| | $P = 8$ | 0.34004 | 0.028727551 | 0.822012797 |